\documentclass[fleqn,usenatbib]{mnras}

\usepackage{newtxtext,newtxmath}
\usepackage[T1]{fontenc}

\DeclareRobustCommand{\VAN}[3]{#2}
\let\VANthebibliography\thebibliography
\def\thebibliography{\DeclareRobustCommand{\VAN}[3]{##3}\VANthebibliography}

\usepackage{graphicx}	% Including figure files
\usepackage{amsmath}	% Advanced maths commands

\newcommand{\dev}[2]{\frac{\text{d} #1}{\text{d} #2}}
\newcommand{\pdev}[2]{\frac{\partial #1}{\partial #2}}

\newcommand{\OmegaP}{\Omega_\mathrm{p}}
\newcommand{\rf}{r_\mathrm{f}}

\title[Boundary Layers and Accretor Structure]{Wave-Mediated Boundary Layers of Accretion Discs: Role of Internal Structure of the Accretor}

\author[S. G. D. Turner, R. R. Rafikov and A. A. Philippov]{
Samuel G. D. Turner,$^{1}$
Roman R. Rafikov$^{1,2}$\thanks{E-mail: rrr@damtp.cam.ac.uk}
and Alexander A. Philippov$^{3}$
\\
$^{1}$Department of Applied Mathematics and Theoretical Physics, University of Cambridge, Wilberforce Road, Cambridge CB3 0WA, UK\\
$^{2}$Institute for Advanced Studies, Einstein Drive, School of Natural Sciences, Princeton, NJ 08540, USA\\
$^{3}$Department of Physics, University of Maryland, College Park, MD 20742, USA}

\date{Accepted XXX. Received YYY; in original form ZZZ}

\pubyear{\the\year{}}

\begin{document}
\label{firstpage}
\pagerange{\pageref{firstpage}--\pageref{lastpage}}
\maketitle

\begin{abstract}
Disc accretion onto astrophysical objects with a material surface proceeds through the boundary layer  (BL) --- a radially narrow region in the inner disc where the incoming gas must slow down its rotation before settling onto the surface of the accretor. Here we numerically study a BL in which the angular momentum transport in the layer is accomplished via the excitation of global acoustic waves. While the earlier studies of such wave-mediated BLs  typically modeled the internal structure of the central object as a globally isothermal sphere with sharply rising density profile, here we explore the effect of other internal density and temperature profiles on the mode operation. We model the inner structure of an accretor as a polytropic sphere, allowing a shallower increase of density and a non-trivial temperature profile inside the object. While the mix of acoustic modes observed in our long-duration (1000 inner orbits long) 2D hydrodynamic simulations is a weak function of the polytropic index $n$ of the accretor's structure, the mass accretion rate and the angular momentum flux across the BL show a clear dependence on $n$, both decreasing in amplitude as $n$ is lowered.  Interestingly, in 2D these transport metrics are better correlated not with $n$ but with a total mass inside the central object contained within the simulation domain.  These results improve our understanding of the wave-mediated BL accretion by quantifying the effect of the inner structure of the accretor on the excitation and propagation of acoustic modes mediating the BL transport.
\end{abstract}

\begin{keywords}
accretion, accretion discs -- hydrodynamics -- waves -- methods: numerical
\end{keywords}

%%%%%%%%%%%%%%%%%%%%%%%%%%%%%%%%%%%%%%%%%%%%%%%%%%
%%%%%%%%%%%%%%%%% BODY OF PAPER %%%%%%%%%%%%%%%%%%

%%%%%%%%%%%%%%%%%%%%%%%%%%%%%%%%%%%%
%%%%%%%%%%%%%%%%%%%%%%%%%%%%%%%%%%%%

\section{Introduction}
\label{sec:intro}

%%%%%%%%%%%%%%%%%%%%%%%%%%%%%%%%%%%%

Accretion discs are found throughout the Universe, from the ones around young stars and planets to those around supermassive black holes at the centres of galaxies. Excepting for when the central object is a black hole, the accretor (which we will hereafter refer to simply as a star) has a physical surface which must interact with the inner edge of the disc. Whenever the stellar magnetic field is sufficiently small, the disc will continue essentially uninterrupted down to the stellar surface \citep{Ghosh&Lamb1978}. This physical setup is expected to be the case for a wide range of objects, including young forming giant planets \citep{Owen2016}, young stars such as FU Orioni objects (FUOrs) \citep[e.g.][]{Popham+1993}, cataclysmic variables (CVs) \citep[e.g.][]{Kippenhahn&Thomas1978, Narayan&Popham1993} and onto weakly magnetised neutron stars in low-mass X-ray binaries \citep[e.g.][]{Inogamov&Sunyaev1999, Inogamov&Sunyaev2010, Gilfanov+2003, Revnivtsev&Gilfanov2006, Philippov+2016}.

In these systems, there will necessarily be a boundary layer (BL) between the fast-rotating inner edge of the disc (which orbits at approximately Keplerian velocity) and the slow-rotating stellar surface. As inflowing material passes through the BL, it must lose its angular momentum in order to slow from an initial Keplerian velocity. Angular momentum transport in the disc is usually attributed to the magnetorotational instability \citep[MRI;][]{Balbus&Hawley1991}, which requires the angular velocity of the flow to be a decreasing function of radius (${\text{d}\Omega/\text{d}r<0}$) in order to operate. This condition cannot hold in the BL, and the inability of the MRI to operate within the BL has been confirmed numerically \citep[e.g.][]{Pessah&Chan2012}. 

Early studies of the BL \citep{Kippenhahn&Thomas1978, Narayan&Popham1993, Popham+1993, Kley&Lin1996, Piro&Bildsten2004, Balsara+2009, Hertfelder+2013, Dong+2021} used a local, $\alpha$-type viscosity prescription \citep{Shakura&Sunyaev1973} to enable angular momentum transport. However, while its use in the disc is well motivated, the lack of a known turbulent mechanism in the BL makes the reliance on a local viscosity ad hoc. In particular, the highly supersonic shear typical for the astrophysical BLs is known to stabilise the Kelvin-Helmholtz instability \citep{Miles1958}, which might otherwise be expected to drive turbulence in the BL. 

Moreover, any local viscosity will inevitably result in the local radiation of energy released in the BL, which can be up to a half of all the energy liberated in the process of accretion \citep{Popham&Narayan1995}. Such a dramatic release of energy from a relatively small region would lead to large temperatures and thus high energy emission, resulting in soft X-ray excesses in CVs and strong UV components in FUOrs \citep{Pringle1977, Popham&Narayan1995}. However, these excesses are generally not observed \citep{Ferland+1982, Kenyon+1989}, a phenomenon known as the `missing boundary layer' problem.  

A very different paradigm for the BL accretion was proposed in  \citet{Belyaev+2012}, who demonstrated that a finite width, supersonic shear layer such as one would expect in a BL is unstable to a sonic instability \citet{Belyaev&Rafikov2012}, which is similar in nature to the Papaloizou-Pringle instability \citep{Drury1980,Drury1985,Papaloizou&Pringle1984, Narayan+1987, Glatzel1988}. The instability excites acoustic waves, which travel into the disc and the star, transporting angular momentum and energy over large distances. The non-local nature of this mechanism naturally explains the missing boundary layer problem as energy from the BL is liberated over a much larger area than in the local models. The presence of these waves and the associated transport of angular momentum has been shown robustly in hydrodynamic \citep{Belyaev+2012, Belyaev+2013a, Hertfelder&Kley2015, Dittmann2021, Dittmann2024, Coleman+2022a, Coleman+2022b, Fu+2023} and magnetohydrodynamic \citep{Belyaev+2013b, Belyaev&Quataert2018} simulations.

Given the diversity of objects that can accrete through the BL, one may wonder whether the inner structure of the accretor ----- its density and temperature profiles --- may affect the excitation and propagation of the BL-excited acoustic modes. Unfortunately, essentially all existing simulations of wave-mediated BLs assumed that the star had a globally isothermal\footnote{Note that generally this does not imply the thermodynamic response to perturbations to be isothermal, see e.g. \citet{Belyaev2017}, \citet{Dittmann2024}.} structure, with density rapidly (exponentially) increasing with depth, which is likely quite different from the structure of real astrophysical accretors. 

In this work, we depart from the common standard and explore the effect of the internal structure of the accretor on the acoustic mode excitation. We do this by modelling the (unperturbed) structure of the accretor to be that of a polytropic sphere, with the variation of the polytropic index $n$ allowing us to control the rates at which the density and temperature vary inside the star and to study the impact on the operation of acoustic modes.  

Our paper is organized as follows. In Section \ref{sec:methods} we describe the methods and setup of our simulations. We provide a brief overview of two of our simulations in Section \ref{sec:overview}, before looking in more detail at one of the accretion bursts in Section \ref{sec:burst}. Sections \ref{sec:mode_mix}, \ref{sec:accretion} and \ref{sec:BL_structure} describe and compare the mix of acoustic modes present, the magnitude of the accretion and the local structure of the BL respectively across all of our simulations. We discuss our results and place them within the wider context in Section \ref{sec:discussion} before concluding in Section \ref{sec:conclusions}.

%%%%%%%%%%%%%%%%%%%%%%%%%%%%%%%%%%%%
%%%%%%%%%%%%%%%%%%%%%%%%%%%%%%%%%%%%

\section{Methodology} 
\label{sec:methods}

%%%%%%%%%%%%%%%%%%%%%%%%%%%%%%%%%%%%

In this paper we perform simulations of the BL between a disc and non-rotating star for a variety of stellar structures. The simulations are hydrodynamical (i.e. there are no magnetic fields) and use locally isothermal\footnote{We implement a locally isothermal setup by solving the energy equation under an adiabatic equation of state with $\gamma=1.00001$. Every time step, a user source term is triggered which removes any change in thermal energy by setting the internal energy equal to the kinetic energy plus the expected thermal energy given by the locally isothermal temperature. This gives the same results as a true locally isothermal simulation, and so we will continue to refer to our simulations as such.} equation of state (EoS) to treat the perturbations on top of the fixed background state defined in Section \ref{sec:phys_setup}. We use the astrophysical code \texttt{Athena++} \citep{Stone+2020} in a vertically integrated cylindrical $(r,\phi)$ setup, which solves the equations of mass and momentum conservation
\begin{align}
    \label{eq:mass_cons}
&    \pdev{\Sigma}{t} + \nabla\cdot(\Sigma\mathbf{v}) = 0 ,
\\
    \label{eq:mom_cons}
  &  \pdev{\left(\Sigma\mathbf{v}\right)}{t}
    + \nabla\cdot(\Sigma\mathbf{v}\mathbf{v} + P\mathbf{I}) = 0 ,
\end{align}
where $\Sigma$ is the surface density, $t$ the simulation time, $\mathbf{v}$ the fluid velocity, $P$ the (vertically integrated) gas pressure and $\mathrm{I}$ the identity tensor. We note that while the concept of a surface density is straightforward within the disc, it is less so within the star. 

The use of a locally isothermal EoS is known to lead to non-trivial outcomes, such as e.g. non-conservation of angular momentum of the density waves propagating in the disc \citep{Miranda2019,Miranda2020a}. To avoid such issues, our EoS becomes globally isothermal in the disc (see Section \ref{sec:disc}) and preserves wave angular momentum flux there. The main reason behind our choice of the locally isothermal EoS in the full domain is its simplicity.

Throughout this paper we will work in code units, defined such that the BL is initially ($t=0$) centred at $r=1$. Further defining $GM=1$ for the star's gravity ensures that the Keplerian velocity at the stellar surface is unity. The orbital period at ${r=1}$ is therefore $2\pi$ and thus we often measure time in multiples of $t/2\pi$.

%%%%%%%%%%%%%%%%%%%%%%%%%%%%%%%%%%%%

\subsection{Physical Setup}
\label{sec:phys_setup}

%%%%%%%%%%%%%%%%%%%%%%%%%%%%%%%%%%%%

Physically, the simulation domain consists of three regions, a star, a BL, and a disc. To define our variables in all of them, we will first define what the relevant variables would be for an isolated star and disc, before joining them smoothly at the BL (around $r=1$).

We will work with temperature in code units as ${T=P/\Sigma=c_s^2}$, omitting numerical factors from the ideal gas law. Note that, since the Keplerian velocity is unity at $r=1$, the Mach number of the flow there is simply $\mathcal{M}=c_s^{-1}$ where $c_s$ is the local sound speed. It is therefore natural to define a fiducial temperature as
\begin{equation}
    \label{eq:T_0}
    T_0 = \mathcal{M}^{-2} ,
\end{equation}
where $\mathcal{M}$ is the Mach number at $r=1$.

%%%%%%%%%%%%%%%%%%%%%%%%%%%%%%%%%%%%

\subsubsection{Disc}
\label{sec:disc}

Given that this work examines the effect of various stellar structures on the behaviour of the BL, the physical setup of the disc is chosen to be similar to previous work for simplicity. Previous work on the topic has used a globally isothermal temperature profile, with $T=T_0$ throughout. We therefore use an isothermal profile throughout the disc, as
\begin{equation}
    \label{eq:T_d}
    T_d(r) = T_0 .
\end{equation}
The initial surface density profile is chosen as 
\begin{equation}
    \label{eq:disc_den}
    \Sigma_d(r) = r^{-1/2} .
\end{equation}
Previous work has used various exponents for the radial dependence, ranging from $-3/2$ \citep[steeply decaying profile,][]{Coleman+2022a} to $0$ \citep[flat profile,][]{Belyaev+2012,Belyaev+2013a,Dittmann2024}. The initial velocity field is chosen to be purely Keplerian, with ${\mathbf{v} = r^{-1/2}\boldsymbol{\hat{\phi}}}$, not adjusted for pressure support, but see Section \ref{sec:sim-dets}.

%%%%%%%%%%%%%%%%%%%%%%%%%%%%%%%%%%%%

\subsubsection{Star}
\label{sec:star}

The star is initialized in hydrostatic equilibrium, without rotational support (${\boldsymbol{v} = 0}$). If this background was modeled with a globally isothermal EoS, with ${T=T_0}$ everywhere, then $\Sigma=\exp{(\mathcal{M}^2(r^{-1}-1))}$. 

In this work, we instead tested various temperature (and thus initial density) profiles. To this effect, we model the stellar background density and temperature structure as corresponding to a polytrope with a polytropic index $n$ \citep[similar to][]{Takasao2025}, i.e. with ${P\propto\Sigma^{1+1/n}}$, or equivalently $T\propto\Sigma^{1/n}$. Very importantly, this polytropic index only defines the background state in the star and is unrelated to the (locally isothermal) EoS used to follow the evolution of various perturbed quantities. With the polytropic background, requiring that ${T=T_0}$ and ${\Sigma=1}$ at the stellar radius ${r=1}$, hydrostatic equilibrium within the gravitational field of a point mass potential\footnote{We are implicitly ignoring the self-gravity of the material in the simulation.} yields the following temperature and density profiles within the star:%
\begin{align}
    \label{eq:T_s}
&    T_s(r) = T_0 \left[1-\frac{\mathcal{M}^2}{1+n}\left(1-\frac{1}{r}\right)\right] ,
\\
    \label{eq:star_den}
&    \Sigma_s(r) = \left[1 - \frac{\mathcal{M}^2}{1+n}\left(1-\frac{1}{r}\right)\right]^n .
\end{align}
We can recover global isothermality by taking the limit ${n\rightarrow\infty}$, in which case eqs. \eqref{eq:T_s} and \eqref{eq:star_den} reduce to ${T_s(r)=T_0}$ and $\Sigma_s=\exp{(\mathcal{M}^2(r^{-1}-1))}$, as expected.

Throughout this work, we will refer to simulation by the value of $n$ used to describe their stellar structure. 

%%%%%%%%%%%%%%%%%%%%%%%%%%%%%%%%%%%%
 
\subsubsection{Boundary Layer}

We have now defined temperature, density, and velocity profiles in the star and the disc. A smooth transition in the profiles of these quantities is required through the BL. This is especially important for the temperature profile, which is held constant throughout the simulation (due to the locally isothermal EoS), whereas the density and velocity profiles are only used as initial conditions.

We choose to smooth the temperature profile as
\begin{equation}
    \label{eq:T}
    T(r) = \left[T_d(r)^k + T_s(r)^k\right]^{1/k} ,
\end{equation}
where $k$ is a constant which defines the width of the transition region. Larger (smaller) values of $k$ give rise to a radially narrower (wider) transition. The value of $k$ is chosen according to a procedure detailed in Appendix \ref{app:k} and is given by the eq. (\ref{eq:k_value}). 

\begin{figure}
    \centering
    \includegraphics[width=\columnwidth]{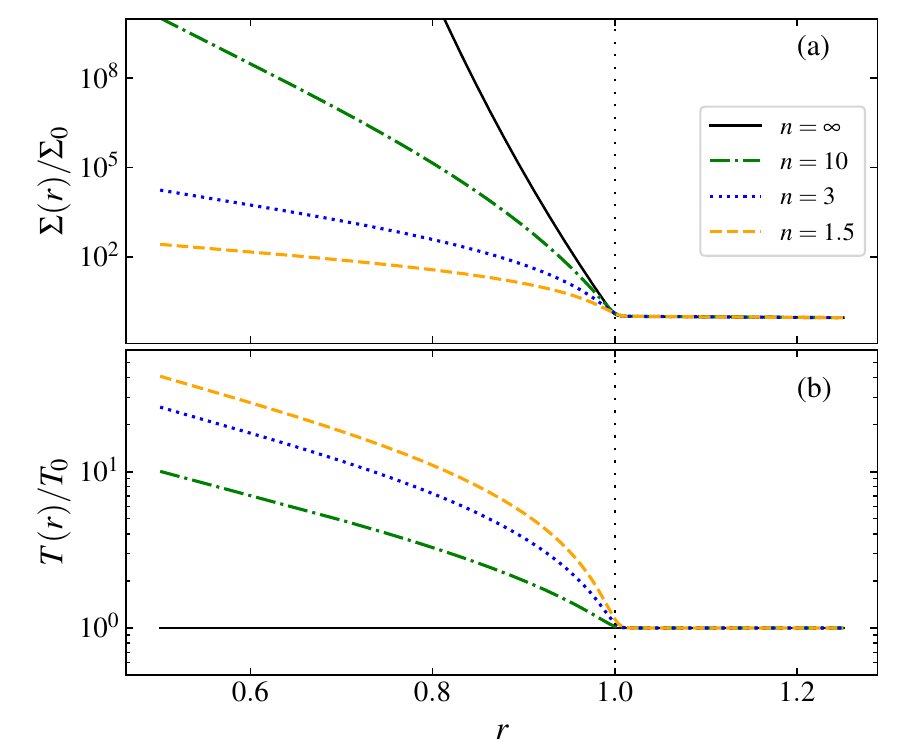}
    \caption{Profiles of (a) surface density $\Sigma(r)$ and (b) temperature $T(r)$ for $\mathcal{M}=10$ defined by equations (\ref{eq:T_s})-(\ref{eq:T}) and used in this work, plotted for several values of the polytropic index of the stellar structure, $n=1.5, 3, 10, \infty$. 
    }
    \label{fig:profiles}
\end{figure}

The smoothing of the density profile is performed in an analogous way to eq. \eqref{eq:T}, with an exponent of $k/n$ to ensure transition of the same width. The initial transition in the azimuthal velocity is performed over a radial width ${\delta r = 0.01}$. In Figure \ref{fig:profiles} we plot the profiles of $\Sigma(r)$ and $T(r)$ for several representative values of $n$.

\begin{table}
    \centering
    \caption{Summary of all the 2D simulations presented in this work. The six columns respectively contain the name of the simulation, the mach number, the polytropic index of the star, the random seed, the inner radius and the resolution of the grid.}
    \begin{tabular}{ccccccc} \hline
        Simulation & $\mathcal{M}$ & $n$ & Seed & $r_\mathrm{in}$ & Resolution & \\ \hline
        N1 & 10 & 1 & a & 0.861 & $4096^2$ \\
        N1.5.a & 10 & 1.5 & a & 0.861 & $4096^2$ \\
        N1.5.b & 10 & 1.5 & b & 0.861 & $4096^2$ \\
        N1.5.c & 10 & 1.5 & c & 0.861 & $4096^2$ \\
        N2.a & 10 & 2 & a & 0.861 & $4096^2$ \\
        N2.b & 10 & 2 & b & 0.861 & $4096^2$ \\
        N2.5 & 10 & 2.5 & a & 0.861 & $4096^2$ \\
        N3.a & 10 & 3 & a & 0.861 & $4096^2$ \\
        N3.b & 10 & 3 & b & 0.861 & $4096^2$ \\
        N3.c & 10 & 3 & c & 0.861 & $4096^2$ \\
        N4 & 10 & 4 & a & 0.861 & $4096^2$ \\
        N5 & 10 & 5 & a & 0.861 & $4096^2$ \\
        N6 & 10 & 6 & a & 0.861 & $4096^2$ \\
        N8 & 10 & 8 & a & 0.861 & $4096^2$ \\
        N10 & 10 & 10 & a & 0.861 & $4096^2$ \\
        N13 & 10 & 13 & a & 0.861 & $4096^2$ \\
        N16 & 10 & 16 & a & 0.861 & $4096^2$ \\
        N20 & 10 & 20 & a & 0.861 & $4096^2$ \\
        Ninf.a & 10 & $\infty$ & a & 0.861 & $4096^2$ \\
        Ninf.b & 10 & $\infty$ & b & 0.861 & $4096^2$ \\
        Ninf.c & 10 & $\infty$ & c & 0.861 & $4096^2$ \\
         &  &  &  &  & \\
        N1.5.r661 & 10 & 1.5 & a & 0.661 & $4096^2$ \\
        N1.5.r288 & 10 & 1.5 & a & 0.288 & $4096^2$ \\
        Ninf.r917 & 10 & $\infty$ & a & 0.917 & $4096^2$ \\
        Ninf.r935 & 10 & $\infty$ & a & 0.935 & $4096^2$ \\
         &  &  &  &  & \\
        M6.N1.5 & 6 & 1.5 & a & 0.691 & $2048^2$ \\
        M6.N3 & 6 & 3 & a & 0.691 & $2048^2$ \\
        M6.N6 & 6 & 6 & a & 0.691 & $2048^2$ \\
        M6.N10 & 6 & 10 & a & 0.691 & $2048^2$ \\
        M6.N20 & 6 & 20 & a & 0.691 & $2048^2$ \\
        M6.Ninf & 6 & $\infty$ & a & 0.691 & $2048^2$ \\ \hline
    \end{tabular}
    \label{tab:all_sims}
\end{table}

%%%%%%%%%%%%%%%%%%%%%%%%%%%%%%%%%%%%

\subsection{Mass and Angular Momentum Transport Characteristics} 
\label{sec:mdot_theory}

%%%%%%%%%%%%%%%%%%%%%%%%%%%%%%%%%%%%

One of our key goals is to quantify the mass and angular momentum transport, to compare the efficiency of the accretion across different values of $n$. The local accretion rate is given as
\begin{equation}
    \label{eq:Mdot}
    \dot{M}(r) = -2\pi r\langle\Sigma v_r\rangle ;
\end{equation}
here and throughout this work, the angled brackets show an azimuthal average as
\begin{equation}
    \label{eq:az_av}
    \langle Q\rangle = (2\pi)^{-1}\int_0^{2\pi} Q\mathrm{d}\phi .
\end{equation}
Along with a mass flux, there is also an angular momentum flux, which can be broken into two contributions, one from advection of angular momentum and the other due to Reynolds stress. For our purposes, it is sufficient to consider the contribution due to the Reynolds stress driven by the acoustic modes, and defined as
\begin{equation}
    \label{eq:C_S}
    C_S = -2\pi r^2\langle\Sigma v_r(v_\phi - v_{\phi,0})\rangle,
\end{equation}
where $v_{\phi,0}$ is a reference angular velocity. In this work we define the reference velocity as the density weighted mean angular velocity,
\begin{equation}
    \label{eq:v_phi0}
    v_{\phi,0} = \frac{\langle\Sigma v_\phi\rangle}{\langle\Sigma\rangle} ,
\end{equation}
\citep[following e.g.][]{Fromang&Nelson2006, Belyaev+2013a} although we should note that other choices, such as ${\langle v_\phi\rangle}$, are also used \citep{Coleman+2022b}.

Since the seminal work of \citet{Shakura&Sunyaev1973}, it is standard practice to quantify transport processes within discs through an $\alpha$ parameter, which relates the local stress to pressure within the disc. Underpinning the $\alpha$ model is the assumption that the transport processes are local phenomena \citep{Balbus&Papaloizou1999}, which is not true of the globally-propagating acoustic waves. Nevertheless, following \citep{Coleman+2022b}, we can define two $\alpha$ parameters, one each for the accretion rate and angular momentum flux as:
\begin{align}
    \label{eq:alpha_acc}
&    \alpha_\mathrm{acc} \equiv \frac{\dot{M}}{2\pi\Sigma c_s^2}\Omega ,
\\
    \label{eq:alpha_stress}
&    \alpha_\mathrm{stress} \equiv \frac{C_S}{2\pi r^2\Sigma c_s^2} .
\end{align}
In the viscous model of \citet{Shakura&Sunyaev1973}, these two quantities are equivalent (and positive) by definition. Note that a positive (negative) value of $\alpha_\mathrm{acc}$ signifies accretion (decretion) and a positive (negative) $\alpha_\mathrm{stress}$ signifies outward (inward) angular momentum transport. In stark contrast to the viscous case, under accretion driven by acoustic modes these two quantities are different not just in magnitude but frequently in sign as well.

%%%%%%%%%%%%%%%%%%%%%%%%%%%%%%%%%%%%

\subsection{Simulation Details}
\label{sec:sim-dets}

%%%%%%%%%%%%%%%%%%%%%%%%%%%%%%%%%%%%

The simulation domain covers a radial range of $[r_\mathrm{in}, 4]$ and $(0,2\pi)$ in $\phi$. The value of $r_\mathrm{in}$ is a function of $\mathcal{M}$ (but not $n$), and is chosen\footnote{Some simulations described in Section \ref{sec:accretion} use smaller values of $r_\mathrm{in}$.} such that $\Sigma(r_\mathrm{in})=10^7$ for a globally isothermal setup. This choice is the same as made in \citet{Coleman+2022a}, and is based on the observation that the width of the BL depends strongly on $\mathcal{M}$ but not on $n$, as will be shown later. The grid spacing in the radial (azimuthal) direction is logarithmic (uniform).

We apply standard periodic boundary conditions in the $\phi$ direction. The radial boundaries use a `do-nothing' condition, where the ghost cells contain values specified by the initial conditions. This ensures that the inner (outer) radial boundary maintains the initial structure of the star (disc). Since we are not supplying a steady influx of mass at the outer boundary, and in the absence of any transport mechanism (e.g. MRI) in the bulk of the disc, our simulations are not designed to reach a steady state.

Each simulation is initialised in 1D (radial only), where it is run for 100 orbits of the inner edge of the disc (i.e. up to ${t/2\pi=100}$) to allow the initial conditions to adjust to the radial pressure support and relax into a steady state. It is then restarted in 2D, where a perturbation is added to the radial velocity in the disc (but not in the star or BL) to seed the acoustic instability. Each grid cell is given a different velocity perturbation according to 
\begin{equation}
    \label{eq:noise}
    \mathrm{v}_r = AX , 
\end{equation}
where $X\in[-1,1]$ and $A=10^{-2}$ is the amplitude of the perturbation. 

The simulation is then run for 1000 orbits (${t/2\pi=1000}$). Note that throughout this work we will refer to durations expressed in orbits as well as ${t/2\pi}$. Unless specified otherwise, orbits refers to the duration of a Keplerian orbit at the inner edge of the disc (i.e. at ${r=1}$), and so an orbit is equivalent to ${t=2\pi}$. Our procedure for outputting different simulation variables is described in Appendix \ref{sec:outputs}.

%%%%%%%%%%%%%%%%%%%%%%%%%%%%%%%%%%%%

\subsection{Parameter Space}

%%%%%%%%%%%%%%%%%%%%%%%%%%%%%%%%%%%%

A list of all the simulations used in this work is shown in Table \ref{tab:all_sims}. Given that we are interested in the effect of the stellar structure, the primary variable of interest is $n$. The majority of the simulations use the same Mach number ${\mathcal{M}=10}$ (in between ${\mathcal{M}=9}$ and $12$ for which a large number of runs were performed in \citealt{Coleman+2022a,Coleman+2022b}), covering a large range of $n$; some of them differ in their value of $r_\mathrm{in}$. Certain combinations are run multiple times with different seeds (i.e. when the simulation is restarted in 2D, different random numbers are used to seed the acoustic instability). A smaller number of simulations are performed with ${\mathcal{M}=6}$, again covering multiple values of $n$. Note that the simulations with $n=\infty$ are globally isothermal, and therefore allow for comparison with previous work.

%%%%%%%%%%%%%%%%%%%%%%%%%%%%%%%%%%%%
%%%%%%%%%%%%%%%%%%%%%%%%%%%%%%%%%%%%

\section{Overview of individual simulations} 
\label{sec:overview}

%%%%%%%%%%%%%%%%%%%%%%%%%%%%%%%%%%%%

Before presenting our main results, it is informative to describe in some detail the evolution of a couple of our simulations for different $n$ and compare their results. To simplify the interpretation of the perturbation patterns that we emerge in the course of these runs, we start by reminding the reader the morphologies of the modes typically excited in the BL.  

%%%%%%%%%%%%%%%%%%%%%%%%%%%%%%%%%%%%

\subsection{Typical Mode Morphologies}
\label{sec:theory}

%%%%%%%%%%%%%%%%%%%%%%%%%%%%%%%%%%%%

\begin{figure}
    \centering
    \includegraphics[width=\columnwidth]{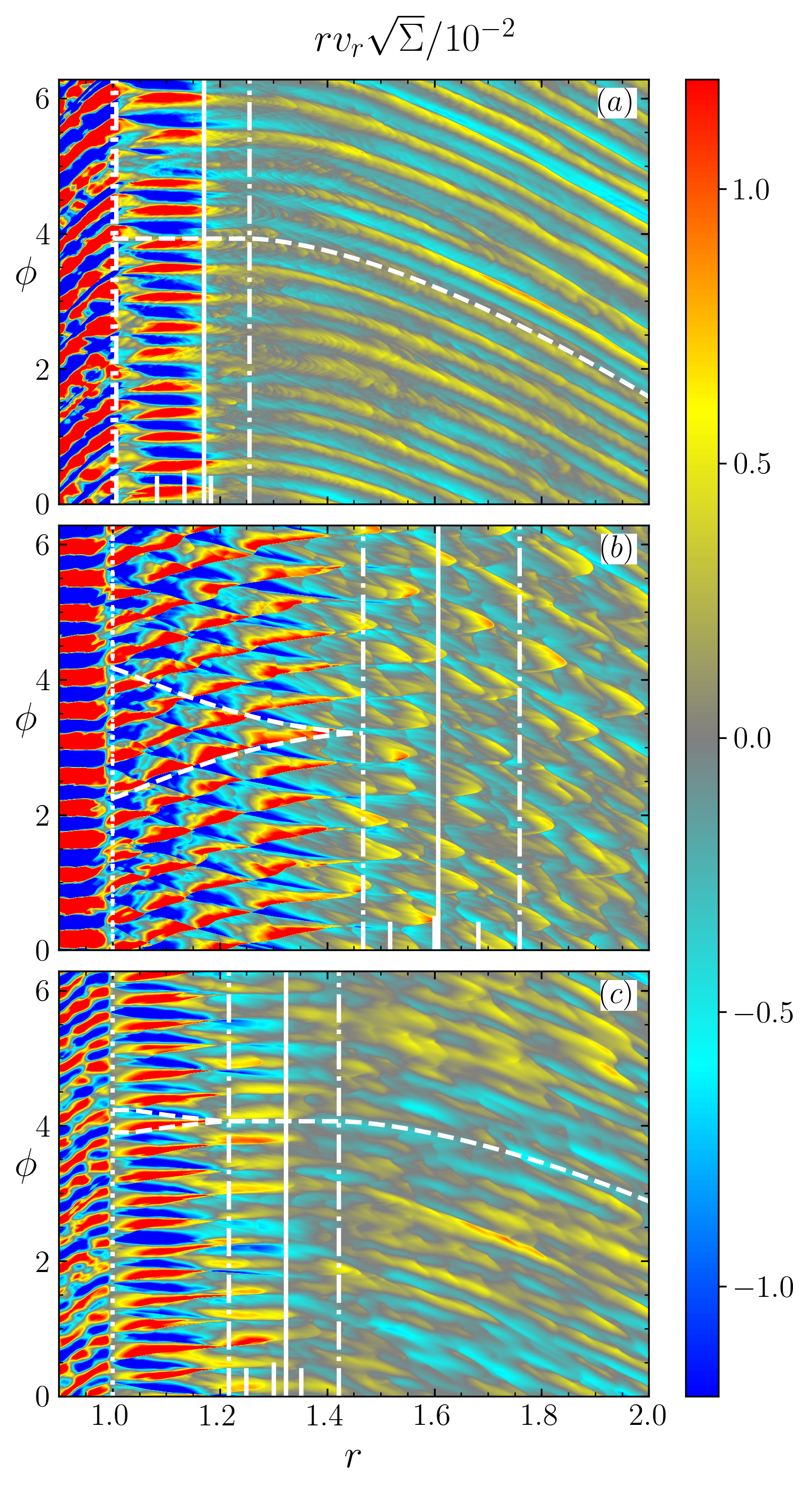}
    \caption{Snapshots of ${rv_r\sqrt{\Sigma}}$, which is a proxy for the wave action, in the $(r,\phi)$ plane. (a): A typical upper mode from the globally isothermal simulation Ninf.b at ${t/2\pi=95}$ when ${\Omega_\mathrm{max}=0.880}$. This mode has ${m=15}$, ${\OmegaP = 0.828}$. (b): A typical lower mode from the globally isothermal simulation Ninf.a at ${t/2\pi=650}$ (${\Omega_\mathrm{max}}=0.849$). This mode has ${m=13}$, ${\OmegaP = 0.494}$. (c): A mixed mode from the globally isothermal simulation Ninf.a at ${t/2\pi=150}$ when ${\Omega_\mathrm{max}}=0.852$. This mode has ${m=17}$, ${\OmegaP = 0.674}$. In each panel, the vertical dotted line shows the location of $r=1$.  The dashed line shows the predicted shape of the wavefront, calculated using eq. \eqref{eq:WKB} and \eqref{eq:wavefront}, with $m$ and $\OmegaP$ measured directly from the data. The vertical solid line is the corotation radius, and the short vertical line at the bottom of each figure the Keplerian value (i.e. the corotation radius if ${\Omega=\Omega_K}$. The vertical dot-dashed lines are the limits of the region where $k_r^2<0$. The corresponding short lines at the bottom of each panel are the Keplerian Lindblad Resonances.
    }
    \label{fig:mode_overlay}
\end{figure}

\citet{Belyaev&Rafikov2012} identified three different branches of solutions for the acoustic modes excited in the BL, termed \textit{upper}, \textit{middle} and \textit{lower} . However, as was first found by \citet{Belyaev+2013a}, simulations typically only show \textit{upper} and \textit{lower} modes, and we will not discuss the \textit{middle} branch further in this work. The \textit{upper} and \textit{lower} branches have very different morphologies and dispersion relations, as we will see next. In Figure \ref{fig:mode_overlay} we illustrate typical mode morphologies by plotting the quantity ${rv_r\sqrt{\Sigma}}$, which should be conserved for acoustic modes preserving their angular momentum flux as they propagate \citep{Belyaev+2013a}.

Before discussing the morphology of individual modes, it is helpful to consider how acoustic modes propagate through the disc. For a given azimuthal wavenumber $m$ (or $k_\phi=m/r$) and pattern speed $\OmegaP$, the radial wavenumber $k_r(r)$ can be found according to \citet{Belyaev+2013a}: 
\begin{equation}
    \label{eq:WKB}
    m^2\left(\Omega - \OmegaP\right)^2
    = c_s^2\left( k_r^2 + \left(\frac{m}{r}\right)^2 \right) + \kappa^2,
\end{equation}
where $\Omega(r)$ and $\kappa(r)$ are the angular and radial epicyclic frequencies respectively. The first term on the right hand side of eq. (\ref{eq:WKB}) is simply $c_s^2k^2$ where ${k=\left(k_r^2+k_\phi^2\right)^{1/2}}$ is the mode wavenumber. In the limit of a tightly wound spiral ${|k_r/k_\phi|\gg1}$ and eq. \eqref{eq:WKB} becomes the standard WKB dispersion relation for waves in discs \citep{Goldreich&Tremaine1978}. The shape of the wavefronts of the modes is given by
\begin{equation}
    \label{eq:wavefront}
    \dev{r}{\phi} = \pm\frac{rk_\phi}{k_r} = \pm\frac{m}{k_r} .
\end{equation}

At the corotation radius (${\Omega=\OmegaP}$), the left hand side of eq. \eqref{eq:WKB} vanishes, and so $k_r^2<0$. This imaginary $k_r$ must extend over a certain radial range, over which the wavefront is radial (since the real part of $k_r$ is zero) and the mode evanescent.

Assuming that $\Omega$ and $\kappa$ are Keplerian, and taking the limit of a tightly wound spiral, the limits of the evanescent region where ${k_r^2<0}$ are the radii of the Inner and Outer Lindblad Resonances
\begin{equation}
    \label{eq:Lindblad}
    r = r_\mathrm{cor,K}\left(1\pm\frac{1}{m}\right)^{2/3}
\end{equation}
where $r_\mathrm{cor,K}$ is the corotation radius assuming a Keplerian profile (i.e.  ${\OmegaP=\Omega_K(r_\mathrm{cor,K})}$). The presence of the ${(m/r)^2}$ term in eq. \eqref{eq:WKB} means that the true evanescent region is somewhat wider than the space between the Lindblad Resonances.

The location of the evanescent region is determined primarily by the value of $\OmegaP$. Assuming $\OmegaP$ lies between the extreme values of ${\Omega(r)}$, there must be two corotation radii, one in the BL and one in the disc (see Figure \ref{fig:delta_defs} for an example of a typical ${\Omega(r)}$). The evanescent region inside the BL is always very narrow, and we will thus ignore it to focus on the one in the disc. 

We illustrate these statements by showing an example on an $m=15$ upper mode from our $n=\infty$ simulation Ninf.b in Figure \ref{fig:mode_overlay}a. The values of ${\OmegaP = 0.828}$ and the maximum of $\Omega(r)$,  ${\Omega_\mathrm{max}=0.880}$, are close, and the resulting evanescent region (demarcated by the vertical dot-dashed lines) is broad and encompasses the BL. Upper modes have ${k_r\neq0}$ within the star and $k_r=0$ just outside the BL, which agrees with the mode pattern in Figure \ref{fig:mode_overlay}a. Further out in the disc, where $k_r$ becomes real, the modes gets wrapped up by the differential rotation. The white dashed line shows the analytic prediction from eq. \eqref{eq:WKB}, and is a good match for the shape of the mode pattern.

In contrast to the upper modes, lower modes are characterised by ${k_r=0}$ within the star and ${k_r\neq0}$ in the disc due to their lower pattern speed. An $m=13$ lower mode is shown in Figure \ref{fig:mode_overlay}b, which has a low ${\OmegaP = 0.494}$ and ${\Omega_\mathrm{max}}=0.849$. Wavefronts launched from the BL into the disc have a positive gradient, shown by the lower part of the dashed line. As the wave propagates further into the disc, $k_r$ reduces and the wavefront bends towards the normal to the BL. This continues until ${k_r=0}$ at the inner dot-dashed line. Here,  the mode is strongly reflected (it is evanescent around the corotation radius, which sits far in the disc due to the low $\OmegaP$), and the returning wavefront (upper part of the dashed line) overlaps with the outgoing wavefront, creating the characteristic criss-cross pattern in the inner disc, forming a resonant cavity. Just as for the upper mode, the analytic prediction given by the white dashed line is a good match.

Sometimes, due to the non-trivial $\Omega(r)$ structure, we encounter modes that appear to transition between the upper and lower branches, see Figure \ref{fig:mode_overlay}c. Here we show an ${m=17}$ mode with ${\OmegaP = 0.674}$ and ${\Omega_\mathrm{max}}=0.852$. Within the star, the $k_r\neq0$ shape is consistent with an upper mode, but the inner disc region appears to show a trapped structure, reminiscent of a lower mode. Fundamentally, this is an upper mode, but the lower pattern speed (c.f. ${\OmegaP = 0.828}$ in the top panel of Figure \ref{fig:mode_overlay}) means that the evanescent region does not contain the BL. Instead, $k_r\neq0$ within the disc (although note ${k_\phi/k_r\gg1}$). The white dashed line showing the analytic prediction shows the path of reflected and transmitted components. The observed wave action suggests that the majority of the wave is reflected, there is evidence in the outer disc of spirals consistent with a transmitted component.

In edition to the upper and lowers modes, there are two other modes identified by \citet{Coleman+2022a} which are worth noting. The first of these are resonant modes, which typically have relatively low values of $m$. Similarly to the lower modes, these modes are trapped in the resonant cavity in the inner region of the disc. However, unlike the lower modes, they have a very different dispersion relation.

There are also the so-called vortex-driven modes. Vortices, forming in the inner disc due to the Rossby wave instability \citep{Lovelace+1999, Li+2000, Li+2001, Johnson&Gammie2005, Ono+2016, Ono+2018, Coleman+2022a}, excite spiral waves that propagate out through the disc (similar to the upper modes). These vortices, and thus their spiral arms, are found to merge over time, often culminating in a single vortex producing a one-armed spiral \citep{Ono+2018, Coleman+2022a}.

Finally, it is important to note that the presence of a single, pure mode is a rare occurrence within our (and previous work's) simulations. Much more common is the presence of multiple modes at any given time.

\begin{figure}
    \centering
    \includegraphics[width=\columnwidth]{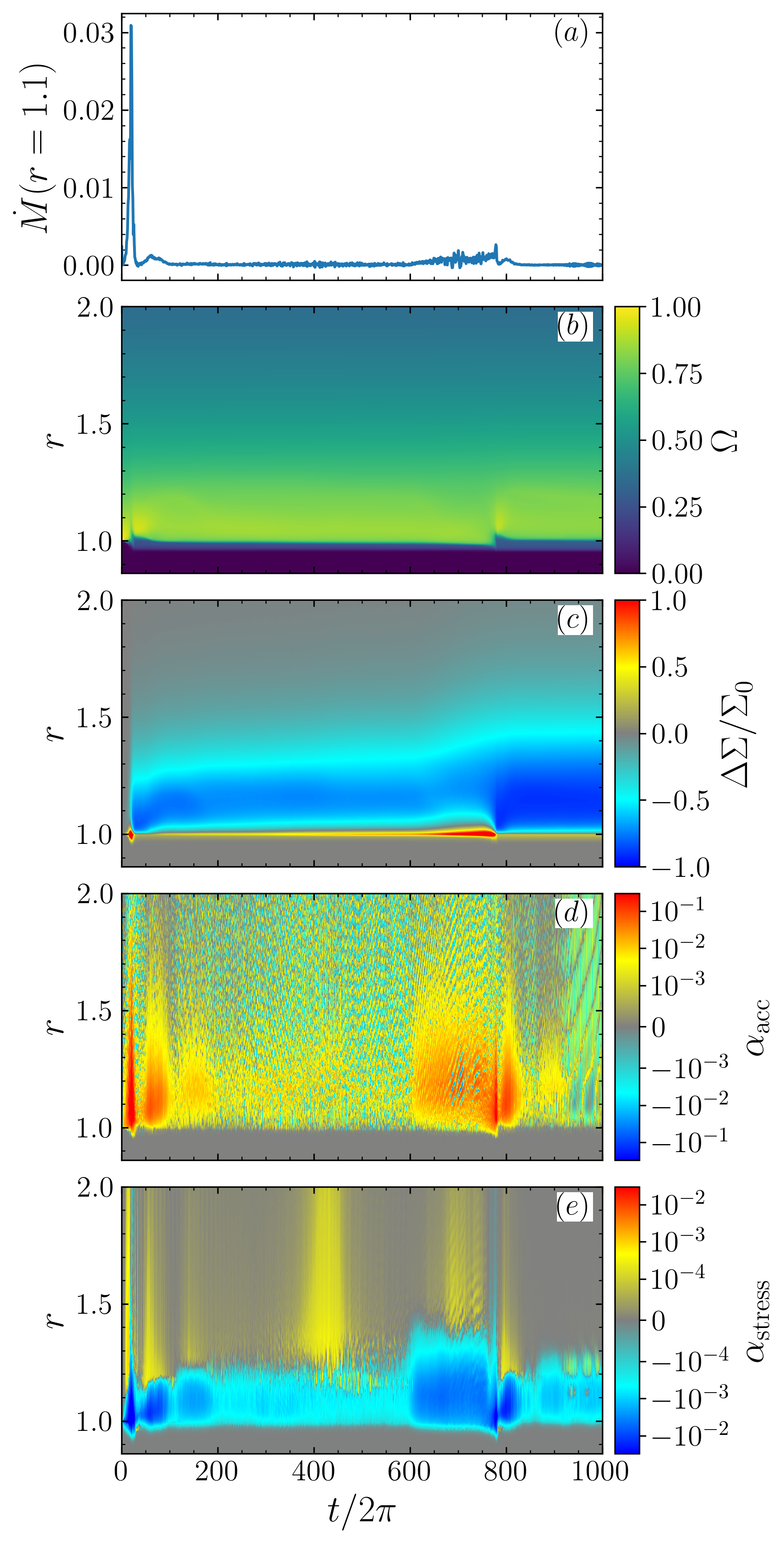}
    \caption{Temporal and radial evolution of various quantities for the ${\mathcal{M}=10}$ globally isothermal simulation Ninf.a. All panels show azimuthally averaged quantities. (a): The local accretion rate at $r=1.1$ (b): The angular velocity (c): The fractional change in the surface density, compared to that at ${t=0}$ (d): $\alpha_\mathrm{stress}$ as defined in eq. \eqref{eq:alpha_stress} (e): $\alpha_\mathrm{acc}$ as defined in eq. \eqref{eq:alpha_acc}. Panels (a), (d) and (e) are smoothed temporally with a box function of width ${t/2\pi=5}$.}
    \label{fig:Ninf_profiles}
\end{figure}

\begin{figure*}
    \centering
    \includegraphics[width=0.9\textwidth]{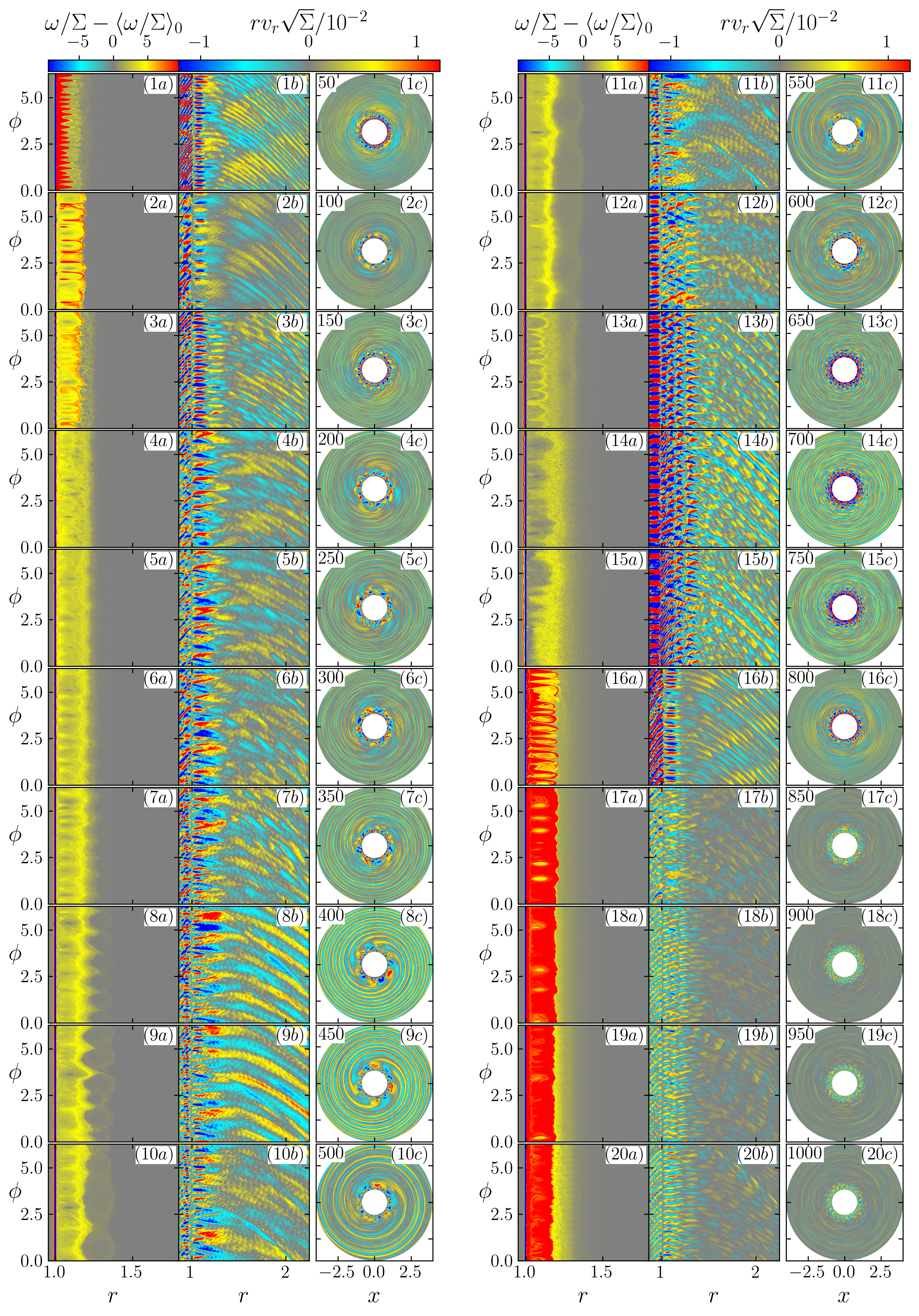}
    \caption{Summary snapshots for a $\mathcal{M}=10$ simulation (Ninf.a) with a $n=\infty$ (globally isothermal) star. Snapshots are shown at intervals $t/2\pi=50$, at times indicated in the upper left corner of the (c) panel. At each time, Panel (a) shows the change in vortensity compared to $t=0$. Panels (b) and (c) show the wave action $rv_r\sqrt{\Sigma}$. Panel (b) shows the inner region $r<2.25$ in the $(r,\phi)$ plane while Panel (c) shows the entire disc. See Section \ref{sec:ninf} for details.}
    \label{fig:Ninf_snaps}
\end{figure*}

%%%%%%%%%%%%%%%%%%%%%%%%%%%%%%%%%%%%

\subsection{Globally isothermal ($n=\infty$) star run}
\label{sec:ninf}

%%%%%%%%%%%%%%%%%%%%%%%%%%%%%%%%%%%%

We now describe two of our ${\mathcal{M}=10}$ simulations in some detail. The first is globally isothermal (Ninf.a) with $n=\infty$ which is similar to those run in previous studies. The second is an ${n=1.5}$ simulation (N1.5.a) described in Section \ref{sec:n1.5}. See Table \ref{tab:all_sims} for details of the parameters used in these simulations. While the stochastic nature of the simulations means we must be careful not to over interpret the results of this comparison, it is useful to provide an idea of the type of behaviour we see. 

Figure \ref{fig:Ninf_profiles} shows the temporal and spatial evolution of various parameters for the $n=\infty$ simulation, in which the density inside the star rises very steeply (exponentially) while $T$ stays constant. We can see that the evolution of the system proceeds in an inherently bursty, stochastic process. Panel (a), which shows the accretion rate at $r=1.1$ (a radius chosen as it lies within the region of the inner disc where the acoustic modes are most active), shows two separate accretion events, one occurring within the first ${t/2\pi=100}$ and the other between ${t/2\pi=600}$ and 800. Associated with these events are large changes to the angular velocity and surface density profiles, which are shown in Panels (b) and (c). Panels (d) and (e) show the effective $\alpha$ parameters for the accretion and angular momentum transport, as defined in eqs. \eqref{eq:alpha_acc} and \eqref{eq:alpha_stress}. These show that, at least within the inner disc, we predominantly find that ${\alpha_\mathrm{acc}>0}$ and ${\alpha_\mathrm{stress}<0}$. The positive value of ${\alpha_\mathrm{acc}}$ indicates a positive accretion rate and thus material moving through the disc and on to the star. The negative ${\alpha_\mathrm{stress}}$ shows that angular momentum is being transported inwards, which is typical of lower (and resonant) modes, whereas for upper (and vortex driven) modes we would expect the opposite signs for both $\alpha$ values \citep{Coleman+2022b}. We will return to this point shortly as we discuss the wave mix seen in these simulations.

We now turn to Figure \ref{fig:Ninf_snaps} showing snapshots of the vortensity $\omega/\Sigma$, where $\omega=|\nabla\times\mathbf{v}|$, and wave action evenly spaced by ${t/2\pi=50}$. We can see that the mix of waves present in the simulation domain is constantly in flux. Upon initialization of the simulation the modes develop quickly, during the first ${t/2\pi=50}$, but they change on a timescale of several orbits. At ${t/2\pi=50}$ a high-$m$ upper mode appears, shown in Panel (1b). The azimuthal wave number of this mode is interesting, as examination of Panel (1b) (and the vortensity structure in (1a)) suggests $m=23$. However, Fourier analysis (detailed in Section \ref{sec:mode_mix}) shows the dominant modes are $m=22$ and $m=24$.

It is during the initial period of ${t/2\pi=50}$ that we see the largest burst of accretion in Panel (1a). It is natural to question whether this is due to our initial conditions, rather than the acoustic modes. We are confident that this is not the case for three reasons. Firstly, our simulations were run for a period of ${t/2\pi=100}$ in 1D, before being restarted in 2D (the restart being defined as ${t=0}$). This means that any transient feature due to our initial conditions not being in hydrostatic equilibrium should have died out during this period. Secondly, a zoomed-in plot of $\dot M(t)$ shows a small initial value that decays to zero within ${t/2\pi=2}$. The increase in the initial burst seen in Panel (a) of Figure \ref{fig:Ninf_profiles} does not start until around ${t/2\pi=10}$. Third, we ran simulations with lower amplitude initial perturbations (as defined in eq. \eqref{eq:noise}), that were otherwise identical to Ninf.a in every way. These simulations also showed initial bursts of accretion, but those bursts were delayed compared to the standard simulation. If the burst is due to a transient feature from the initial conditions, we would expect the burst to occur at the same time regardless of the amplitude of the noise. The observed delay is, however, perfectly consistent with the burst being due to wave action, as an initially smaller amplitude seed perturbation will require a longer time to develop into a wave capable of driving the accretion. Finally, we note that such initial bursts have been routinely observed in other BL studies \citep[e.g.][]{Belyaev+2013a,Coleman+2022b} and are believed to be related to the wave activity while the inner disc has not yet been depleted.

Within this initial burst ${t/2\pi<50}$, Figure \ref{fig:Ninf_profiles} shows a substantial evolution. The profile of $\Omega(r)$ in Panel (b) shows a change in the location of the BL (defined as where $\Omega$ changes from approximately unity to near zero), although the location quickly returns to where it was initially. Panel (c) shows that a large gap is quickly carved out of the $\Sigma(r)$ profile within $r\lesssim1.2$, with this material being dumped on to the star. We remind that because our simulations have no mechanism for mass replenishment in the inner disc, they do not reach a steady state with radially-constant $\dot M$. Panel (d) shows that $\alpha_\mathrm{acc}$ is large and positive throughout this time, as would be expected given the large burst of accretion. Similarly, Panel (e) shows a negative $\alpha_\mathrm{stress}$, demonstrating that the angular momentum is being transported inwards with the material. This does not contradict the typical characteristics of transport by the upper mode present at ${t/2\pi=50}$ since the mode is evanescent close to stellar radius $r=1$. Outside the region $r\gtrsim 1.2$, $\alpha_\mathrm{stress}$ is indeed positive as the upper mode is propagating freely. The important role of the $\partial\Omega/\partial t$ term in the balance of angular momentum \citep{Belyaev+2013a,Coleman+2022b,Cordwell2024} during the initial stages of evolution should also be considered.

Beyond ${t/2\pi=50}$, upper modes dominate for the next 100 orbits, with a particularly clear ${m=17}$ upper mode seen in Panel (4b). There is then a period of relative quiescence, before we see an $m=6$ vortex driven mode at ${t/2\pi=400}$. Six vortices can be seen in Panel (8a), and the associated six spiral arms in Panels (8b-c). The origin of this vortex structure can be seen in Panel (6-7a). As expected, these vortices combine and reduce in number, which can be seen in Panels (9-11a). The merging of the spiral features can be seen most clearly in Panels (10-11c), where the spiral structure only partially fills the disc, in contrast with Panel (8c). Panel (e) of Figure \ref{fig:Ninf_profiles} shows the expected positive values of $\alpha_\mathrm{stress}$ in the outer disc when the spirals are strongest from ${t/2\pi=400}$ to 450.

Starting at ${t/2\pi=600}$ in Panel (12b) of Figure \ref{fig:Ninf_snaps}, we see the development of an $m=13$ lower mode. This mode persists for at least 200 orbits and is still present, although starting to disappear, in Panel (15b). This mode, present clearly in both the star and the disc, is associated with the burst around $t/2\pi\approx 770$ --- large increase in the magnitude of the $\alpha$ parameters in Panels (d) and (e) of Figure \ref{fig:Ninf_profiles}, and also an increase in the accretion rate seen in Panel (a) and a major re-arrangement of $\Sigma$ in the inner disc (Panel c). We describe this burst of accretion in more details in Section \ref{sec:burst}. There is also the suggestion of vortex driven spiral arms around ${t/2\pi=700}$ to 750 in Panels (14-15b), which correspond to a positive $\alpha_\mathrm{stress}$ in Panel (e) of Figure \ref{fig:Ninf_profiles}.

The final ${t/2\pi=200}$ or the simulation, shown in Rows (17-20) in Figure \ref{fig:Ninf_snaps}, shows a relatively quiescent state. Throughout, there appears to be a relatively high $m$ upper mode dominating in the star ($k_r\neq0$) and a lower mode of similar $m$ in the disc (also $k_r\neq0$). In additional, Rows (19-20) show a vortex mode. Panels (19b-c) and (20b-c) show a weak one-armed spiral in the disc. The vortex driving this can be seen in the bottom left corner (${\phi=0}$) of Panel (19a). In Panel (20a), the same vortex appears to have weakened, but the top half of the vortex can be same in a similar location to Panel (19a).

%%%%%%%%%%%%%%%%%%%%%%%%%%%%%%%%%%%%

\subsection{Polytropic ($n=1.5$) star run}
\label{sec:n1.5}

%%%%%%%%%%%%%%%%%%%%%%%%%%%%%%%%%%%%

\begin{figure}
    \centering
    \includegraphics[width=\columnwidth]{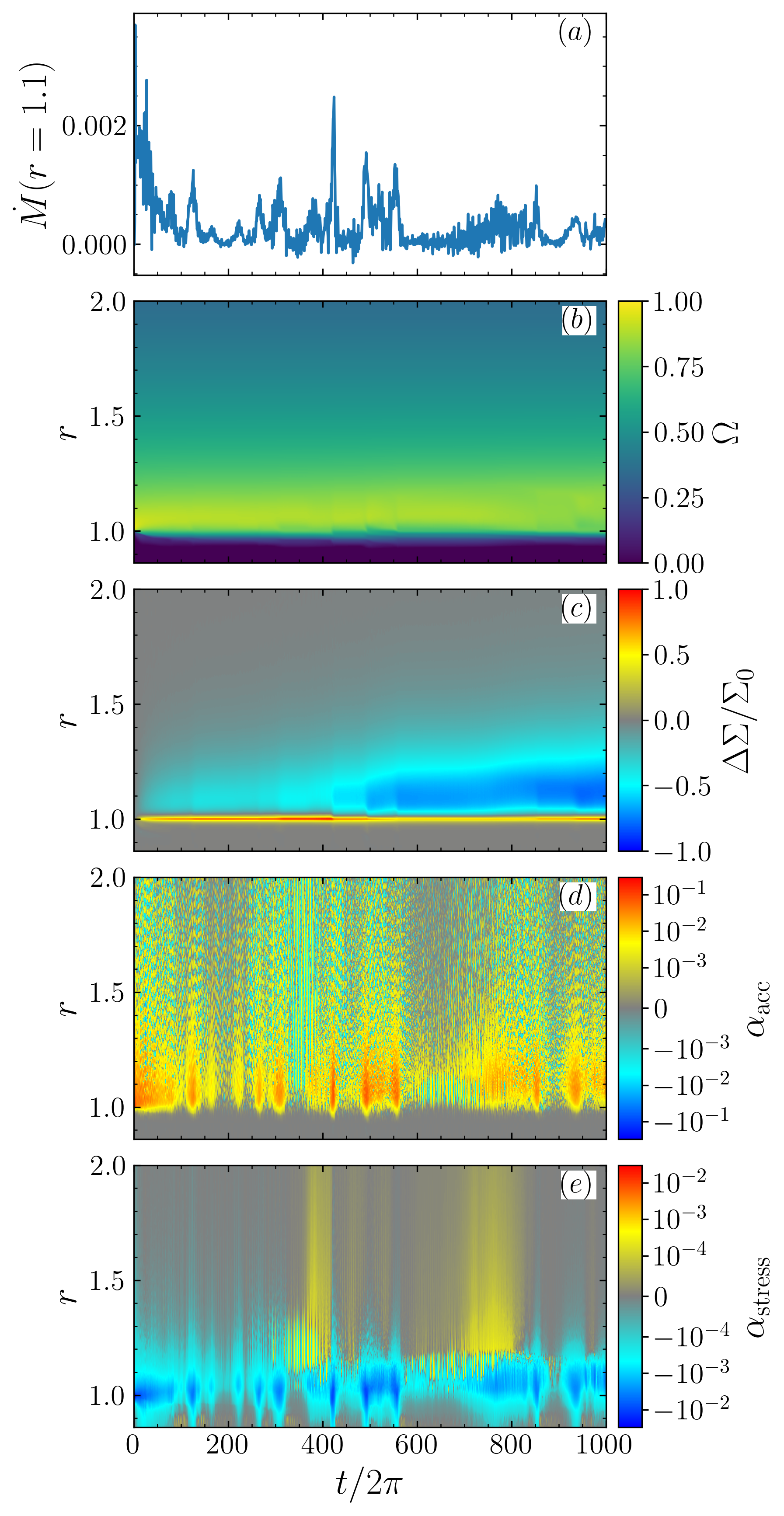}
    \caption{Same as Figure \ref{fig:Ninf_profiles} but for the $n=1.5$ simulation N1.5.a.}
    \label{fig:N1_5_profiles}
\end{figure}

\begin{figure*}
    \centering
    \includegraphics[width=0.9\textwidth]{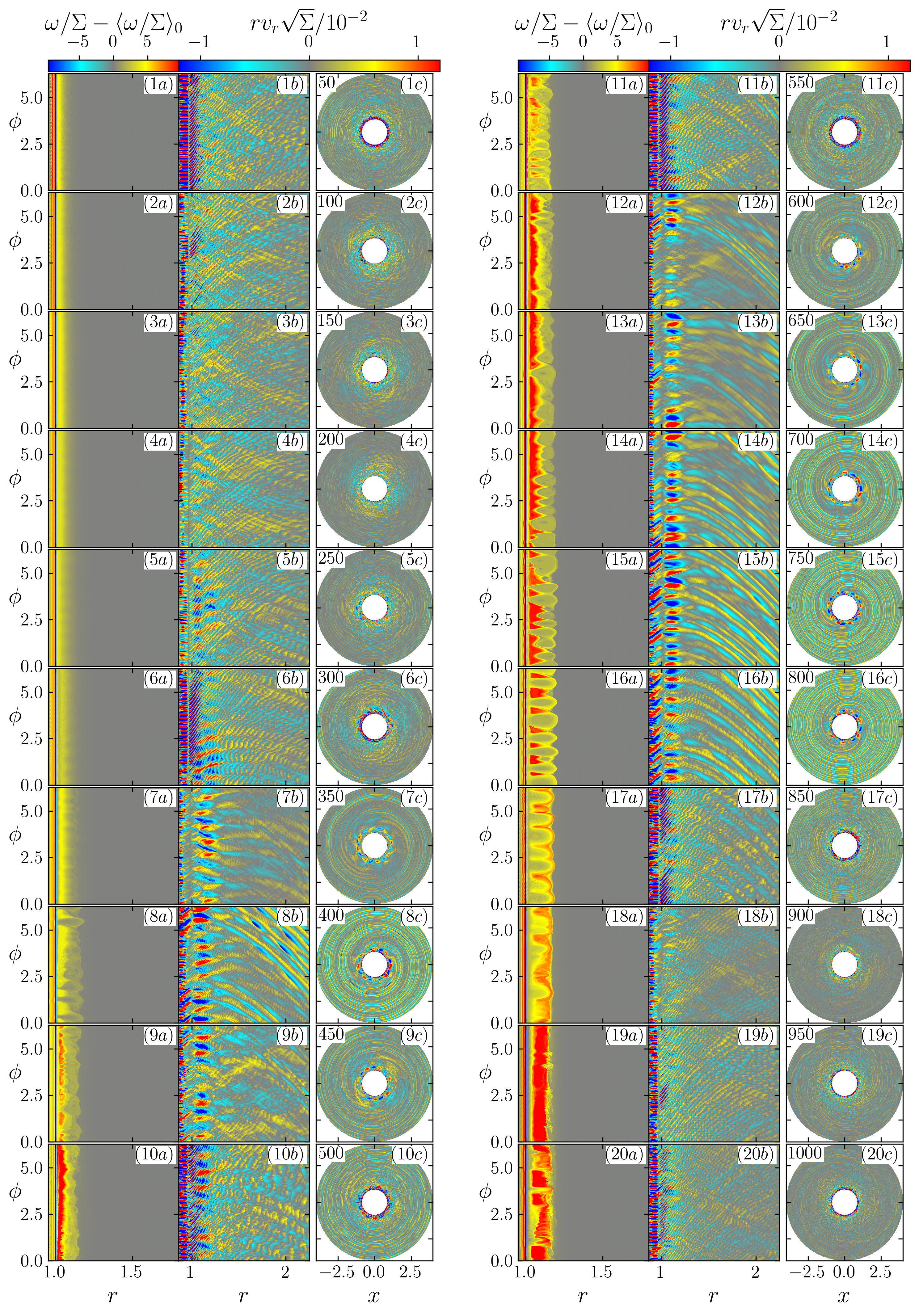}
    \caption{Same as Figure \ref{fig:Ninf_snaps} but for the $n=1.5$ simulation N1.5.a.  See Section \ref{sec:n1.5} for details.}
    \label{fig:N1_5_snaps}
\end{figure*}

We now look at Figures \ref{fig:N1_5_profiles} and \ref{fig:N1_5_snaps}, illustrating the $n=1.5$ simulation N1.5.a. For this stiff stellar structure profile, the temperature inside the star increases towards the centre, see Section \ref{sec:star}. The density, while also increasing, does so much slower than in the isothermal case covered earlier. Comparing Figure \ref{fig:N1_5_profiles}a to the corresponding panel in Figure \ref{fig:Ninf_profiles}, we can immediately see that, while the accretion rate in the $n=1.5$ simulation is bursty, the bursts are weaker and are more frequent than for the globally isothermal simulation in Figure \ref{fig:Ninf_profiles}. This difference is found in other panels as well. The $\Omega$ profile (Panel (b)) shows much less variability in Figure \ref{fig:N1_5_profiles} than Figure \ref{fig:Ninf_profiles}, and the gap in the surface density is considerably shallower and takes significantly longer to clear. The $\alpha$ values, shown in Panels (d) and (e), are also bursty on a shorter timescale in Figure \ref{fig:N1_5_profiles}.

We now look at Figure \ref{fig:N1_5_snaps} are the mode mix present in this simulation. As for the globally isothermal simulation, the modes start to develop quickly, and by ${t/2\pi=10}$ are already present. In this case, the initial ${t/2\pi=50}$ is dominated by lower modes, with no upper modes present. At ${t/2\pi=50}$ shown in Panel (1b), there is a clear lower mode present, although the disc does not exhibit the criss-cross pattern that might be expected. By eye, this mode appears to have ${m=30}$ although Fourier analysis shows a broad peak of power centred at ${m=31}$. This suggests that the mode is not particularly pure, and looking closely at Panel (1b), we can see that the wavelength of the mode within the star does appear to change with $\phi$.

The next period of evolution is relatively quiescent. Panels (2-5b) all show relatively little activity within the disc. Despite this, there are still some ${k_r=0}$ modes present in the star, although their strength is a strong function of $\phi$. In Panel (6b) at ${t/2\pi=300}$, the lower mode returns to being present in the disc.

At ${t/2\pi=350}$, an ${m=8}$ resonant mode is present in the disc, shown in Panel (7b). This persists in some form until ${t/2\pi=450}$. In addition, there is a section of upper mode which is present at ${t/2\pi=400}$. This mode has a limited azimuthal range, and can be seen in Panel (8b) at values of $\phi$ around 0/$2\pi$. The spiral behaviour in the disc is matched by clear ${k_r=0}$ feature in the star, suggesting that the mode is global in the sense that it is present in both the star and disc, despite the limited azimuthal range. Throughout this period from ${t/2\pi=350}$ to 450, there is also a high $m$, ${k_r=0}$ feature present in the star.

From ${t/2\pi=500}$ onwards, the behaviour stays generally similar to that at earlier times, albeit the evolution is different due to the stochastic behaviour. We can pick out time when a global lower mode dominates in Rows (10, 11, 17, 19 and 20); times when the disc is quiescent but there are still high $m$, ${k_r=0}$ features in the star in Row (18); and times when there is an upper mode present, in the disc and star, but there is still the high $m$, ${k_r=0}$ feature in the star in Rows (12-16).

As we saw for the globally isothermal simulation, the presence of the upper modes in the simulation aligns closely with where $\alpha_\mathrm{stress}$ is positive in the outer disc, seen in Panel (e) of Figure \ref{fig:N1_5_profiles}. As in Figure \ref{fig:Ninf_profiles}, the inner disc still features a negative $\alpha_\mathrm{stress}$.

%%%%%%%%%%%%%%%%%%%%%%%%%%%%%%%%%%%%

\subsection{Comparison of the Two Simulations}

%%%%%%%%%%%%%%%%%%%%%%%%%%%%%%%%%%%%

During the preceding discussion, we have made some comparisons between the two simulations we have analysed, but it is worth briefly reiterating the two key differences that appear. Firstly, the accretion rate and associated disc evolution for the globally isothermal simulation (shown in Figure \ref{fig:Ninf_profiles}) shows significantly fewer yet more significant bursts than for the ${n=1.5}$ simulation shown in \ref{fig:N1_5_profiles}. This is discussed more extensively in Section \ref{sec:accretion}.

The second difference is in the mode mix. The $n=1.5$ simulation showed the almost constant presence of high $m$, ${k_r=0}$ features in the star. We interpret these as lower modes that are restricted to the star. In contrast, no such high $m$ features were seen in the globally isothermal simulation. We discuss the mode mix across all simulation in Section \ref{sec:mode_mix}.

%%%%%%%%%%%%%%%%%%%%%%%%%%%%%%%%%%%%
%%%%%%%%%%%%%%%%%%%%%%%%%%%%%%%%%%%%

\section{Anatomy of an Accretion Burst} 
\label{sec:burst}

%%%%%%%%%%%%%%%%%%%%%%%%%%%%%%%%%%%%

As seen in Section \ref{sec:overview}, accretion through the BL is an inherently bursty process, with mass transport onto the star often dominated by a small number of short-duration episodes of greatly increased $\dot M$. The important role of the bursts was recognized previously \citep{Belyaev+2012,Hertfelder&Kley2015}, which motivates us to explore their nature in more details. 

\begin{figure*}
    \centering
    \includegraphics[width=\textwidth]{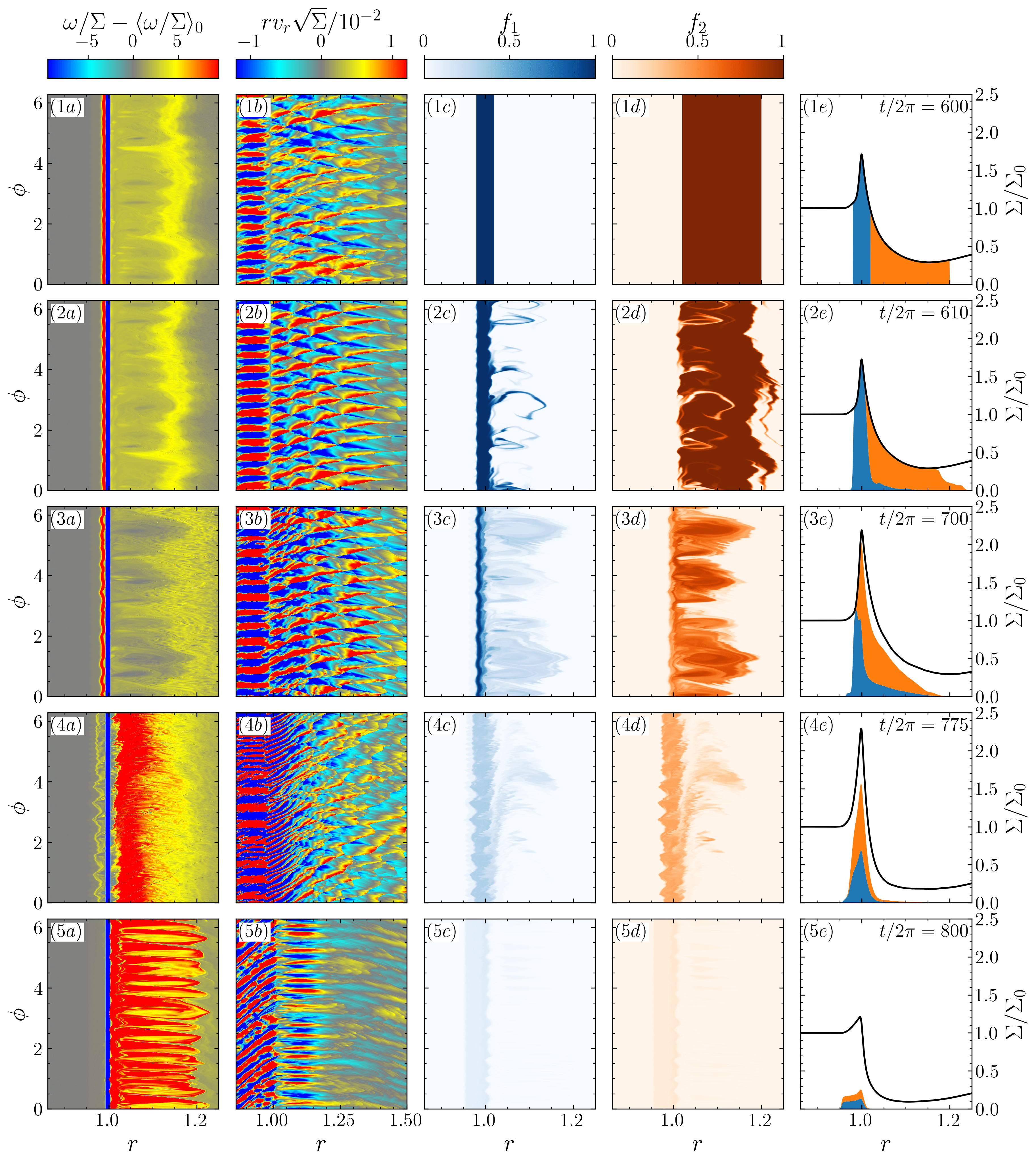}
    \caption{Evolution of the passive scalar fields from ${t/2\pi=600}$ to $800$, following the restart of simulation M10.Ninf.a as detailed in Section \ref{sec:burst}. Each row of the simulation shows the state at ${t/2\pi=600}$, $610$, $700$, $775$ and $800$ respectively. Columns (a) and (b) show the vortensity and wave action in the inner region of the disc. For more detail on these quantities, see Figure \ref{fig:Ninf_snaps} and associated discussion. Columns (c) and (d) show the distribution of the two passive scalar fields respectively. The quantity plotted is the fraction of the surface density at each location which is made up of the passive scalars, and so lies in the range ${[0,1]}$ by definition. The solid black line in Column (e) shows the total surface density divided by the surface density at ${t=0}$. The reason for this choice is to highlight changes in the surface density as a result of the evolution. The blue and orange shaded regions show the fraction of the surface density made up of the scalar fields.}
    \label{fig:passive_scalar}
\end{figure*}

We do this by examining the accretion burst around $t/2\pi\approx 770$ in our $n=\infty$ run Ninf.a, see Section \ref{sec:ninf}. We take a restart from ${t/2\pi=600}$\footnote{As it is not possible to restart a simulation in \texttt{Athena++} while changing the number of passive scalars, the restart is performed using the hdf5 output at ${t/2\pi=600}$ as the input to a new simulation which includes the passive scalars. The hdf5 outputs are saved to double precision (i.e. the same precision as the inbuilt restart files), and so no loss of precision results in this restart method.} in that run (shown in Panels (16a-c) in Figure \ref{fig:Ninf_snaps}), adding two passive scalar fields to track the evolution of mass in the disc. These scalars act as dyes within the flow, and initially flag all the density lying within ${0.98\leq r<1.02}$ and ${1.02\leq r<1.2}$ respectively, see Panel (1e) in Figure \ref{fig:passive_scalar}. This restarted simulation is then run for 200 orbits up to a final time of ${t/2\pi=600}$.

Before looking in detail at the evolution of these passive scalars, we should briefly consider the evolution of the total surface density profile. The solid black line in the right-hand column (column e) of Figure \ref{fig:passive_scalar} shows the surface density divided by the surface density at ${t=0}$. Two features of these plots are immediately obvious. Firstly, the surface density in the inner disc (${1\lesssim r\lesssim 1.5}$) is significantly depleted compared with the initial conditions. This was seen in Figure \ref{fig:Ninf_profiles} and discussed in Section \ref{sec:overview}. The second feature is the enhanced density in the BL (${r\sim1}$) itself, which was not observed in Figure \ref{fig:Ninf_profiles} due to the steep density gradient inside the star.

Before considering the origin of this enhancement, we first consider the evolution of the two passive scalars. The top row of Figure \ref{fig:passive_scalar} shows the initial distribution of the scalar fields at ${t/2\pi=600}$. In Panel (1e) we can see that the first scalar field (blue) follows material that is initially in region of enhanced surface density across the BL, whereas the second scalar field (orange) corresponds to material initially in the inner, depleted region of the disc. While the 2D maps shown in Columns (c) and (d) of Figure \ref{fig:passive_scalar} are informative, it is useful to define some integrated quantities in order to better understand the passive scalars' temporal evolution.

Firstly, we can consider the degree of mixing between the two passive scalar fields as the system evolves. We can quantify this by calculating an overlap integral, defined for two functions $f(r,\phi)$ and $g(r,\phi)$ as
\begin{equation}
    \label{eq:int_2D}
    I_\mathrm{2D} = \frac{\iint fg \mathrm{d}A}{\sqrt{\iint f^2 \mathrm{d}A \iint g^2\mathrm{d}A}}
\end{equation}
where the double integrals are performed over the whole domain and ${\mathrm{d}A\equiv r\mathrm{d}r\mathrm{d}\phi}$. Given that the two scalars are everywhere positive, this quantity lies in $[0,1]$, where values near 0 indicate almost no overlap between the two fields and values near 1 indicate that the two distributions are nearly identical. This quantity is equally dependent on radial and azimuthal distribution. However, as seen in Column (e) of Figure \ref{fig:passive_scalar}, we often care about the radial distribution only. We can therefore calculate a 1D version of eq. \eqref{eq:int_2D}, using azimuthally averaged values of the scalar fields as follows
\begin{equation}
    \label{eq:int_1D}
    I_\mathrm{1D} = \frac{\iint \langle f\rangle\langle g\rangle r\mathrm{d}r}{\sqrt{\iint \langle f\rangle^2 r\mathrm{d}r \iint \langle g\rangle^2r\mathrm{d}r}},
\end{equation}
where ${\langle Q\rangle \equiv 1/2\pi \int Q \mathrm{d}\phi}$. It is straightforward to see that $I_\mathrm{1D}\geq I_\mathrm{2D}$, with equivalence only when the azimuthal distribution of the two passive scalar fields is the same.

The second integrated quantity that is useful to define is the mean radius of each passive scalar. This is defined straightforwardly as
\begin{equation}
    \label{eq:mean_r}
    \bar{r}_i = \frac{\iint\Sigma_i r \mathrm{d}A}{\iint\Sigma_i \mathrm{d}A},
\end{equation}
where $\bar{r}_i$ is the mean radius and $\Sigma_i$ is the surface density of the $i$th passive scalar field. As usual, the integrals cover the whole domain. 

\begin{figure}
    \centering
    \includegraphics[width=\columnwidth]{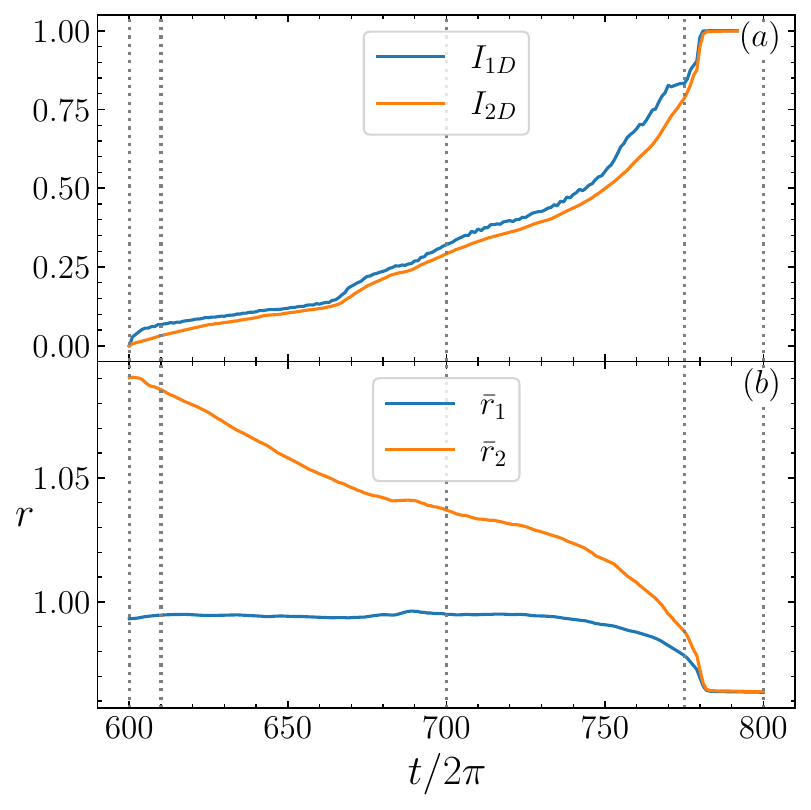}
    \caption{Evolution of various parameters for the passive scalar fields from the restart of simulation M10.Ninf.a. (a): The  overlap integrals as detailed in eqs. \eqref{eq:int_2D} and \eqref{eq:int_1D}. (b): The mean radius of the individual passive scalar fields, as calculated in eq. \eqref{eq:mean_r}. The vertical dotted lines in both panels correspond to the times of the snapshots in Figure \ref{fig:passive_scalar}. }
    \label{fig:passive_scalar_int}
\end{figure}

The time evolution of both the overlap integrals (top panel) and the mean radii (bottom panel) are shown in Figure \ref{fig:passive_scalar_int}. To understand this plot, we can consider Row 2 (${t/2\pi=610}$) of Figure \ref{fig:passive_scalar}. Here, we can see the start of the initial mixing between the two scalar fields. This is proceeding through tendrils of material shown in Panels (2c) and (2d), which are caused by the vortices residing at the flat portion of the $\Omega(r)$ profile in the vicinity of the BL. This initial mixing gives rise to overlap in the 1D profiles (Panel (2e)), even when there is negligible small-scale mixing in the 2D plane. We can therefore consider $I_\mathrm{1D}$ as a measure of the large-scale mixing and $I_\mathrm{2D}$ as an equivalent measure of the small-scale mixing. Returning to Fig \ref{fig:passive_scalar_int}, the apparent co-evolution of the two quantities is clear evidence that the acoustic modes active in the BL at this time naturally lead to mixing on both large and small scales. This can be seen in Row 3 (${t/2\pi=700}$) of Figure \ref{fig:passive_scalar}, where the azimuthal distributions of the two passive scalar fields (Panels (3c) and (3d)) is almost identical. These distributions closely follow the vortensity map shown in Panel (3a), suggesting that vortices play an important role in mixing. Returning to Figure \ref{fig:passive_scalar_int}, we can see that both $I_\mathrm{1D}$ and $I_\mathrm{2D}$ reach essentially unity after only 180 orbits. Since unity implies that the two fields have identical distributions, this shows that the acoustic modes can very quickly lead to a loss of knowledge of initial conditions of material in and around the BL.

We are now in a position to return to consideration of the density enhancement in the BL. Figure \ref{fig:passive_scalar} clearly shows the persistent increase in this enhancement at least over the period from ${t/2\pi=600}$ to 775. as seen in Panels (1-4e). During this enhancement, material from the inner disc (orange passive scalar) is being transported inwards and onto the BL. This is shown in the bottom panel of Figure \ref{fig:passive_scalar_int} as the mean radius of this scalar moves inwards over this period. However, the mean radius of the blue passive scalar, which starts in the BL, is essentially unchanged, at least up to ${t/2\pi=750}$. During this period, a global $m=13$ lower mode is present throughout the star and disc. We can therefore conclude that, at least for this particular burst, the lower mode is able to accrete material from the inner disc, but is unable to move that material into the star itself. As a result, there must be a build up of material in the BL which leads to the seen density enhancement.

That, however, is not the full story, as between ${t/2\pi=775}$ and 800 (Panels (4e) and (5e) respectively), the density enhancement almost completely vanishes. This evolution coincides with a spike in the accretion rate around ${t/2\pi=775}$, which can be seen in Panel (a) of Figure \ref{fig:Ninf_profiles}. Associated with this is a change in the mode morphology present. Panel (4b) clearly shows that, while the ${m=13}$ lower mode persists in the star, a much higher $m$ lower mode now dominates the disc and is also present in the star. This mode is very short lived, as by ${t/2\pi=800}$, Panel (5b) shows an upper mode with no evidence of a lower mode, of any $m$. Around the same time, the bottom panel Figure \ref{fig:passive_scalar_int} shows that the mean radius of both passive scalar fields moves inwards very quickly, before stopping abruptly inside the star at ${t/2\pi=780}$. Although it is unclear why, we suggest that this high $m$ lower mode is able to move material into the star in a way the low $m$ mode could not. In support of this, Panels (d) and (e) of Figure \ref{fig:Ninf_profiles} show that the region of non-zero $\alpha$ reaches significantly further into the star around ${t/2\pi=775}$. It is also not clear why the high $m$ mode is so short lived, although it may be that the rapid evolution of the system (which appears to be due to this mode) brings about its destruction.

While this analysis has been for a single burst, taken from a single simulation, similar behaviour of the metrics such as the evolution of $\Sigma$ and stresses around the burst time is seen in different bursts across different simulations. Thus, we suggest that the preceding discussion applies, at least qualitatively, quite generally.

%%%%%%%%%%%%%%%%%%%%%%%%%%%%%%%%%%%%
%%%%%%%%%%%%%%%%%%%%%%%%%%%%%%%%%%%%

\section{Mode analysis} 
\label{sec:mode_mix}

%%%%%%%%%%%%%%%%%%%%%%%%%%%%%%%%%%%%

\begin{figure*}
    \centering
    \includegraphics[width=0.9\textwidth]{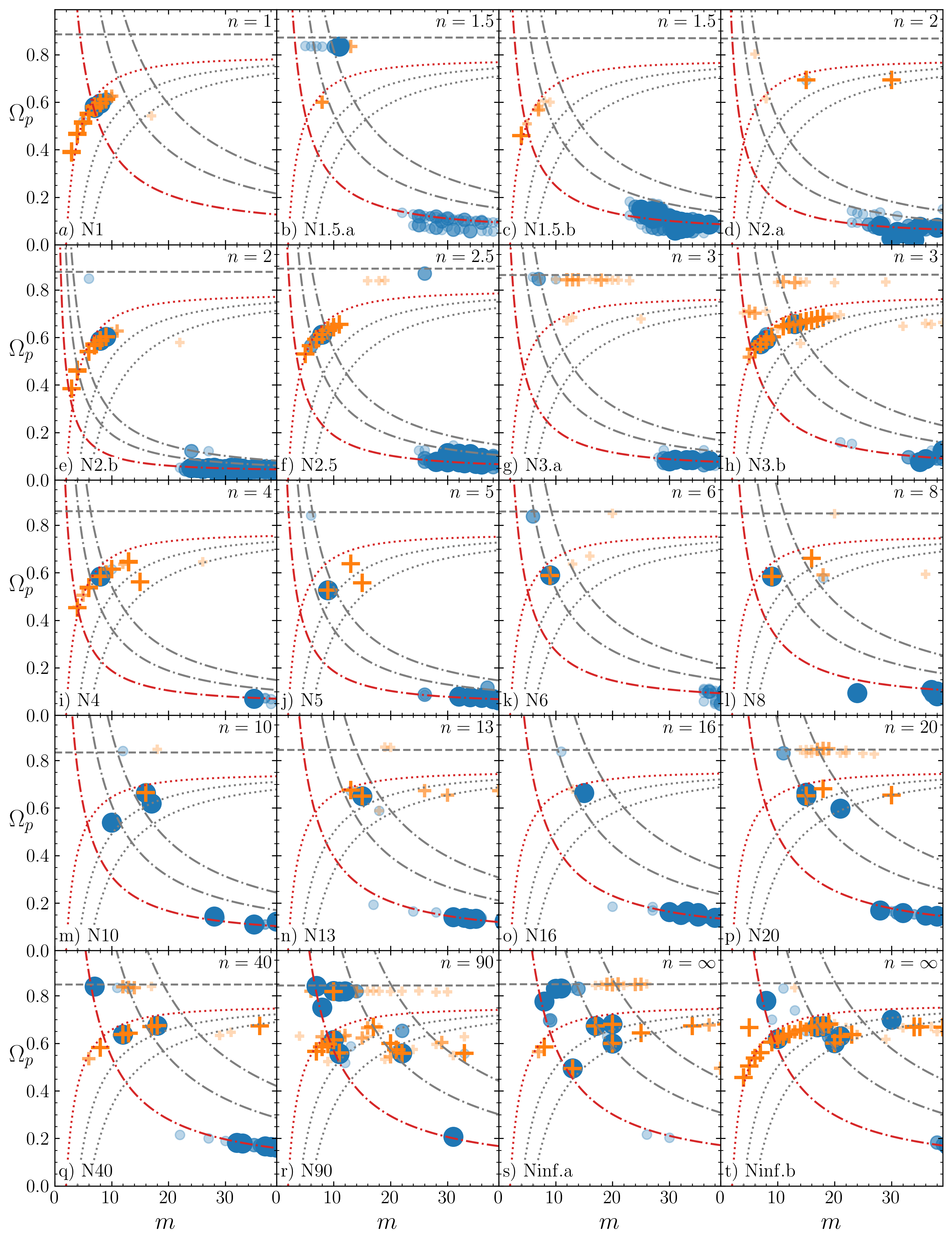}
    \caption{Detected modes from a range of simulations. In each panel, blue circles and orange pluses correspond to modes detected within the star and disc respectively. The red dotted and dot-dashed curves show the dispersion relations for upper and lower modes, respectively, from \citet{Coleman+2022a}, while the grey curves show their second and third harmonics. See text for details. The grey dashed horizontal line shows the value of $\Omega_\mathrm{max}$.}
    \label{fig:all_modes}
\end{figure*}

One conclusion from the discussion of two individual simulations in Section \ref{sec:overview} was that the mode mix appeared very different between the globally isothermal and the $n=1.5$ simulations. Here we extend this comparison across a range of simulations with different $n$. To do this, we need a way to identify the modes present that is more scientific than by eye. For this, we use the high cadence azimuthal profiles of the perturbation variables at $r=0.93$ (which lies halfway between the inner edge of the simulation domain and $r=1$) and $r=1.1$ (from here on, we will refer to these radii as the star and the disc respectively). The details of the mode detection procedure are described in Appendix \ref{sec:mode-detect}.

The detected modes are shown in Figure \ref{fig:all_modes}, with detections from the star shown as blue points and those from the disc as orange pluses. Also, the red dot-dashed line gives the analytic dispersion relation for the lower modes given by eq. \eqref{eq:lower_analytic}. Similarly, the red dotted line shows the dispersion relation for the upper modes given by eq. \eqref{eq:upper_analytic}. Full details of these analytic dispersion relations can be seen in Appendix \ref{app:mode_eqs}. The grey dot-dashed (dotted) lines show the second and third harmonics of the lower (upper) modes. 

Modes that reached the highest detection threshold are shown as larger and less transparent symbols than those at the lower level. The stochastic nature of the simulations is immediately obvious, as panels for simulations with the same input parameters (apart from the initial random seed), such as Panels (g) and (h), display very different modes. While this stochasticity complicates any interpretation of trends, we can draw a few conclusions.

\begin{figure}
    \centering
    \includegraphics[width=\columnwidth]{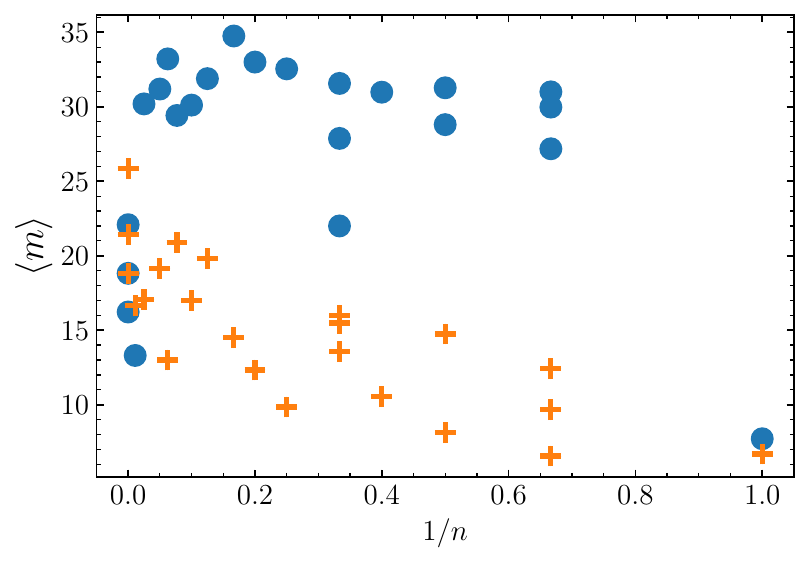}
    \caption{The average azimuthal wavenumber $m$ for the modes detected within the star (blue points) and the disc (orange pluses) for each simulation, plotted against $1/n$.}
    \label{fig:mean_m}
\end{figure}

Firstly, there appears to be a systematic difference between the types of modes detected in the star and the disc. The stellar detections appear to be primarily high $m$ lower modes, whereas those in the disc are mainly low $m$ upper modes. Figure \ref{fig:mean_m} shows the average value of $m$ of the modes detected in the star and disc, for each simulation. From this, it is clear that the disc modes have systematically lower azimuthal wave numbers than those in the star.

In Section \ref{sec:overview}, we noted that the $n=1.5$ simulation (N1.5.a) featured high $m$ lower modes within the star at almost all times. These modes can be seen in the lower-right corner of Panel (b) of Figure \ref{fig:all_modes}. There appear to be detections for all values of ${m\gtrsim25}$, and also a significant scatter in the pattern speed, even for modes with the same $m$. Similar features are seen in Panels (c-f), covering $n=1.5$, 2 and 2.5 simulations. As the value of $n$ increases, fewer upper modes are detected, and their pattern speeds become more consistent. However, the value of $\langle m\rangle$ for the stellar modes in Figure \ref{fig:mean_m} remains ${\gtrsim30}$ for the majority of the simulations. These values are likely to be underestimates, affected by our arbitrary limit of $m=40$, above which we do not attempt to detect modes. The main outliers to this trend are N1 (Panel (a) of Figure \ref{fig:all_modes}), where we do not detect any lower modes, and the globally isothermal simulations (Panels (s) and (t)). The reason for this is unclear, but one possibility is that the presence of a radial temperature gradient within the star favours the excitation of high $m$ lower modes.

Turning now to the disc modes in Figure \ref{fig:mean_m}, there does appear to be a trend where simulations with lower values of $n$ typically have modes with lower values of $m$. Looking at the various panels in Figure \ref{fig:all_modes}, there is large variation in the number of disc modes detected. Panels such as (k) and (o) have almost none, in contrast to the plethora seen in Panels such as (h) and (t). Despite this variation, the trend seen in Figure \ref{fig:mean_m} is reasonably tight, suggesting that the observed correlation is reasonably robust.

%%%%%%%%%%%%%%%%%%%%%%%%%%%%%%%%%%%%
%%%%%%%%%%%%%%%%%%%%%%%%%%%%%%%%%%%%

\section{Mass Accretion and Angular Momentum Transport} 
\label{sec:accretion}

%%%%%%%%%%%%%%%%%%%%%%%%%%%%%%%%%%%%

\begin{figure}
    \centering
    \includegraphics[width=\columnwidth]{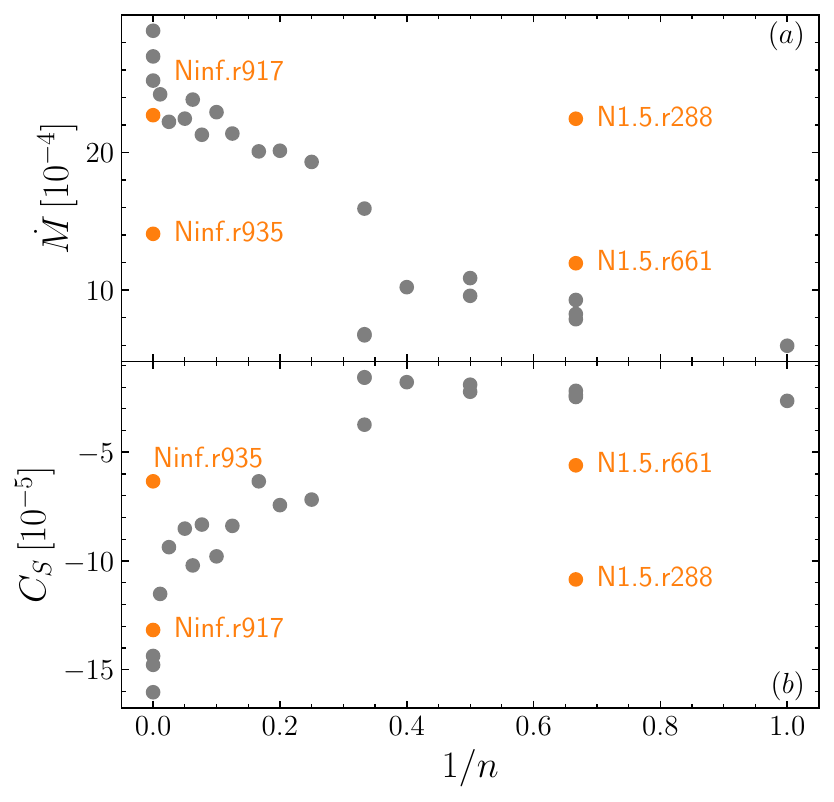}
    \caption{The accretion rate (a) and wave driven angular momentum transport (b) as a function of $1/n$, calculated for the first ${t/2\pi=100}$, for all our $\mathcal{M}=10$ simulations. The grey points correspond to the first group of simulations listed in Table \ref{tab:all_sims}, while the labelled orange points correspond to the second group of runs in the same table, which explore the effect of different inner radii of the domain $r_\mathrm{in}$. }
    \label{fig:Mdot_trends_ninv}
\end{figure}

\begin{figure}
    \centering
    \includegraphics[width=\columnwidth]{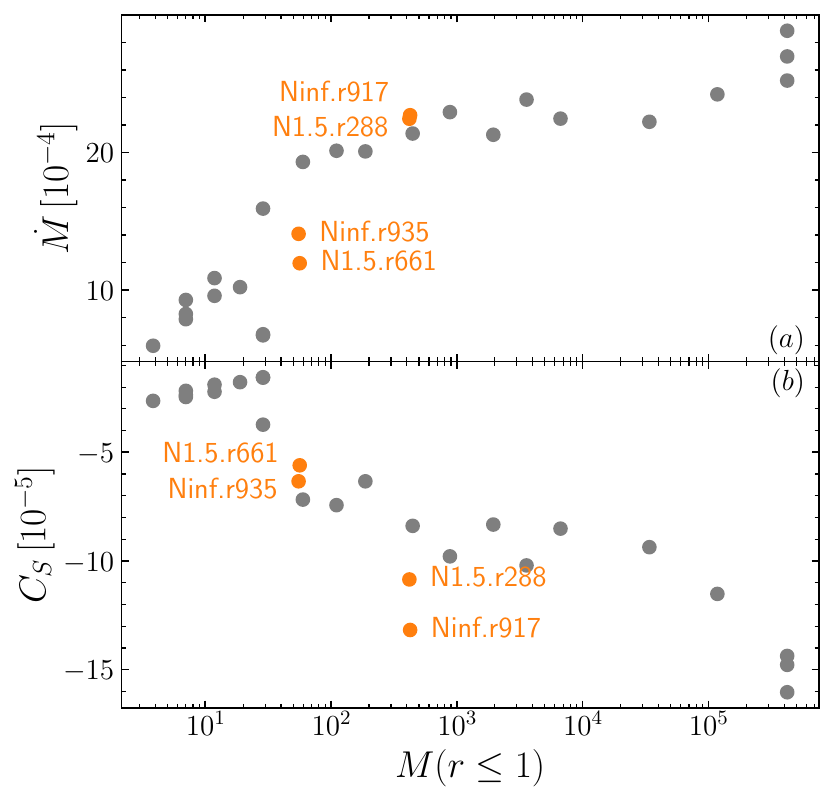}
    \caption{As for Figure \ref{fig:Mdot_trends_ninv}, but the $x$-axis now shows the mass contained within ${r\leq1}$ for each simulation, rather than $1/n$. }
    \label{fig:Mdot_trends_M}
\end{figure}

\begin{figure}
    \centering
    \includegraphics[width=\columnwidth]{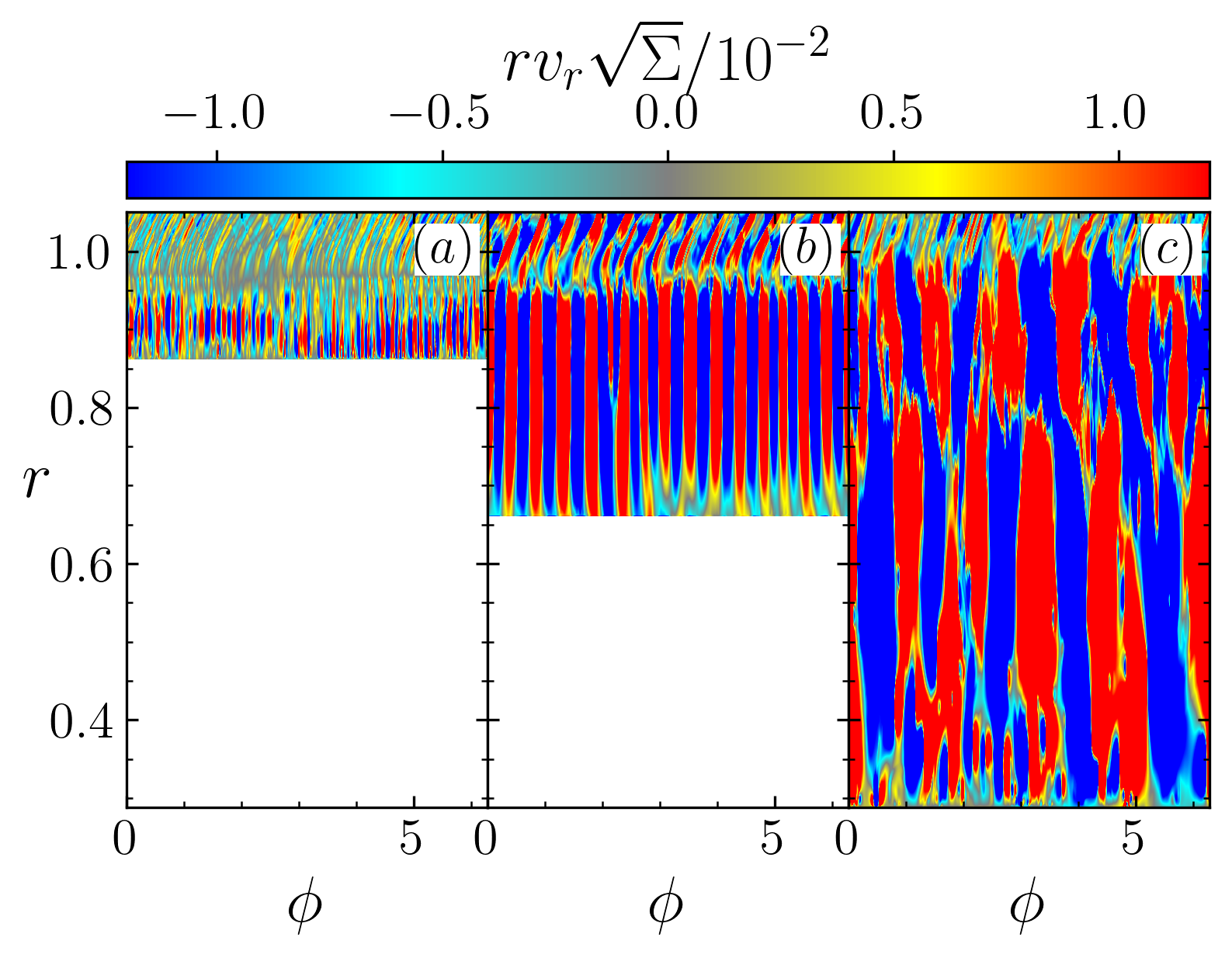}
    \caption{Snapshots showing the wave action in the star and inner edge of the disc for simulation N1.5 (a) N1.5.r661 (b) and N1.5.r288 (c) with $r_\mathrm{in}=0.861$, $0.661$ and $0.288$, respectively. One can see that wave activity penetrates all the way into the star, regardless of the depth of the inner boundary.}
    \label{fig:depth}
\end{figure}

\begin{figure}
    \centering
    \includegraphics[width=\columnwidth]{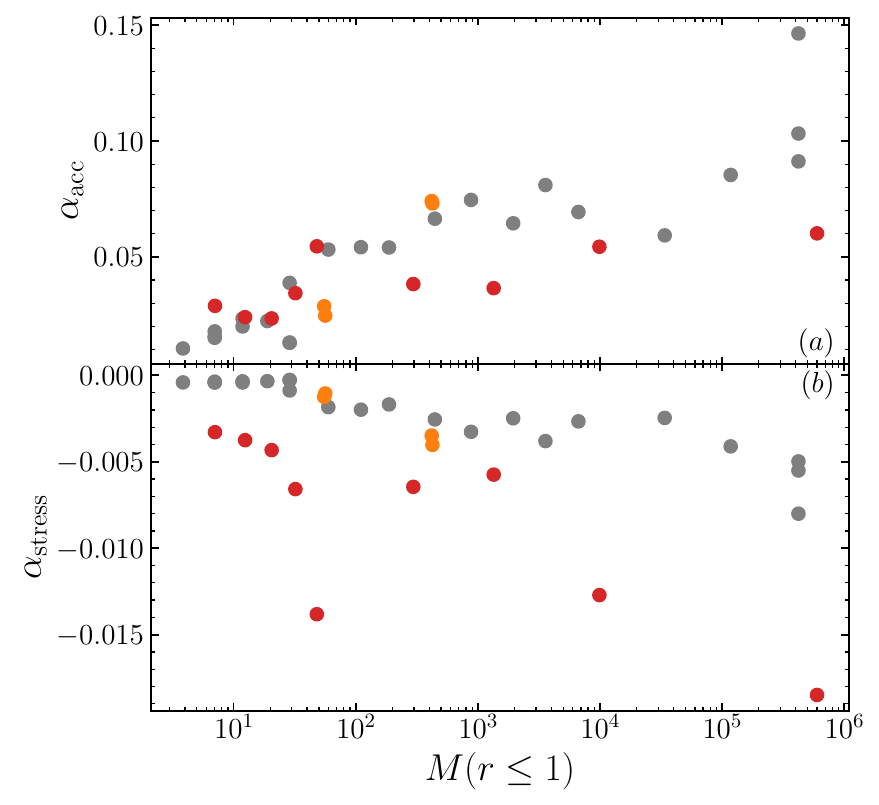}
    \caption{Values of $\alpha_\mathrm{acc}$ (a) and $\alpha_\mathrm{stress}$ (b), as defined in eqs. \eqref{eq:alpha_acc} and \eqref{eq:alpha_stress} respectively, as a function of the mass enclosed within ${r\leq1}$ for each simulation. The values are calculated from the initial ${t/2\pi=100}$. The grey and orange points are as for those in Figure \ref{fig:Mdot_trends_ninv}. The red points are for the ${\mathcal{M}=6}$ simulations for the final group of simulations from Table \ref{tab:all_sims}.  }
    \label{fig:Mdot_trends_M6}
\end{figure}

Having discussed the variation in the modes excited for various polytropic indices, we can now consider whether mass and angular momentum transport also vary with $n$. This comparison is somewhat complicated by two factors. Firstly, the inherently stochastic, bursty nature of the accretion (see Sections \ref{sec:overview} and \ref{sec:burst}) means that there will be a large natural variance in any measurement, simply due to whether or not there is a burst of accretion in the time range considered. Secondly, there is only a finite amount of material in the inner regions of the disc which is available to be accreted in the absence of mass infill from outside. Unlike real systems where the inner region will be replenished by the action of the MRI in the outer disc, no such effect is present in our simulations. Therefore, when considering accretion over very long timescales, we would expect that the average accretion rate will be determined primarily by the amount of material available in the inner disc, not on how efficient the accretion process actually is. 

With these two issues in mind, we choose to focus our comparison on the initial period up to ${t/2\pi=100}$. All our simulations show initial bursts of accretion activity in this interval (starting around ${t/2\pi=10}$). Our logic is that by focusing on the transport processes occurring during the initial burst (inasmuch as it is a true result of the acoustic wave activity), when the inner disc is largely undepleted, we can sample the conditions which would be more typical for a disc in a true accretion steady state, when there is a mechanism of refilling the inner disc by transporting mass from the outer disc regions. Additionally, the initial surface density profile is constant across all simulations. This choice allows for the fairest comparison between all the simulations, although we still expect a large amount of variance for any given simulation. 

Figure \ref{fig:Mdot_trends_ninv} shows the accretion rate $\dot{M}$ and the wave driven angular momentum transport $C_S$, calculated at $r=1.1$ over the period ${t/2\pi=0-100}$. We chose the radius $r=1.1$ as it lies in the middle of the region over which the disc modes are active, although all radii in the inner disc give similar qualitative results. The data are plotted against $1/n$ to allow the globally isothermal simulations with ${n=\infty}$ to be shown on the same scale. The grey points in both panels correspond to the first grouping of simulations in Table \ref{tab:all_sims}. These simulations are all ${\mathcal{M}=10}$ and have identical simulation domains. As such, they differ only in the temperature and density structure of their star. It is clear from both panels of Figure \ref{fig:Mdot_trends_ninv} that there is a clear trend with $n$, with larger polytropic indices giving rise to significantly larger accretion of both mass and angular momentum. It is worth noting explicitly that the negative values of $C_S$ imply that angular momentum is being transported inwards along with mass. This peculiar behaviour is a feature of the evolutionarily dominant lower (and resonant) modes (see Section \ref{sec:mdot_theory} for details).

To understand this result, we first need to investigate whether this effect arises as a result of the local structure of the star (i.e. is dependent primarily on $n$), or on something else which is itself a function of $n$. To do this, we ran four additional simulations, two globally isothermal (${n=\infty}$) and two with ${n=1.5}$. These simulations each have different inner boundaries\footnote{Previously, \citet{Coleman+2022a} discussed the impact of varying $r_\mathrm{in}$ in their globally isothermal ($n=\infty$) simulations.}, which are shallower (deeper) for the globally isothermal (${n=1.5}$) simulations. The precise inner boundaries are shown in the second group in Table \ref{tab:all_sims}. The results from these simulations are shown in the labelled orange points in Figure \ref{fig:Mdot_trends_ninv} and are illustrated in Figure \ref{fig:depth}. It is immediately clear that these new simulations do not fit the previous trend. Looking at the two ${n=1.5}$ results, they are both significantly more active than the three standard simulations. This activity is a function of the location of the inner boundary, with a deeper boundary giving rise to a larger initial burst. The same effect is seen for the globally isothermal simulations, except now the shallower boundaries of the new simulations give rise to smaller initial bursts.

These results imply that it is not the local, near-BL stellar structure that determines the magnitude of the burst, but instead something of a global nature. We therefore recast the plots from Figure \ref{fig:Mdot_trends_ninv} and show both $\dot{M}$ and $C_S$ as a function of the stellar mass (i.e. the mass within $r<1$) in each simulation. These results are shown in Figure \ref{fig:Mdot_trends_M}. Note here that the direction of the $x$-axis has been reversed, as the low $n$ simulations which appeared on the right of Figure \ref{fig:Mdot_trends_ninv} are now on the left of Figure \ref{fig:Mdot_trends_M}. While the agreement is not perfect, the orange points are much closer to the trend of the grey points in Figure \ref{fig:Mdot_trends_M} than they were in \ref{fig:Mdot_trends_ninv}. This is especially true for the top panel, showing $\dot{M}$. 

One may wonder if the dependence of $\dot M$ and $C_S$ on $r_\mathrm{in}$ obvious in Figure \ref{fig:Mdot_trends_ninv} for $n=1.5,\infty$ simply reflects the lack of numerical convergence of our simulations with respect to $r_\mathrm{in}$. However, the correlation with enclosed stellar mass persists also for numerous runs with a fixed $r_\mathrm{in}=0.861$ across the range of $n$, see Table \ref{tab:all_sims}. There is no a priory reason for that to be the case, unless it is the total stellar mass in the simulation domain which determines the magnitude of the accretion burst. This would make this correlation a physical effect rather than a numerical artifact.

\begin{figure*}
    \centering
    \includegraphics[width=\textwidth]{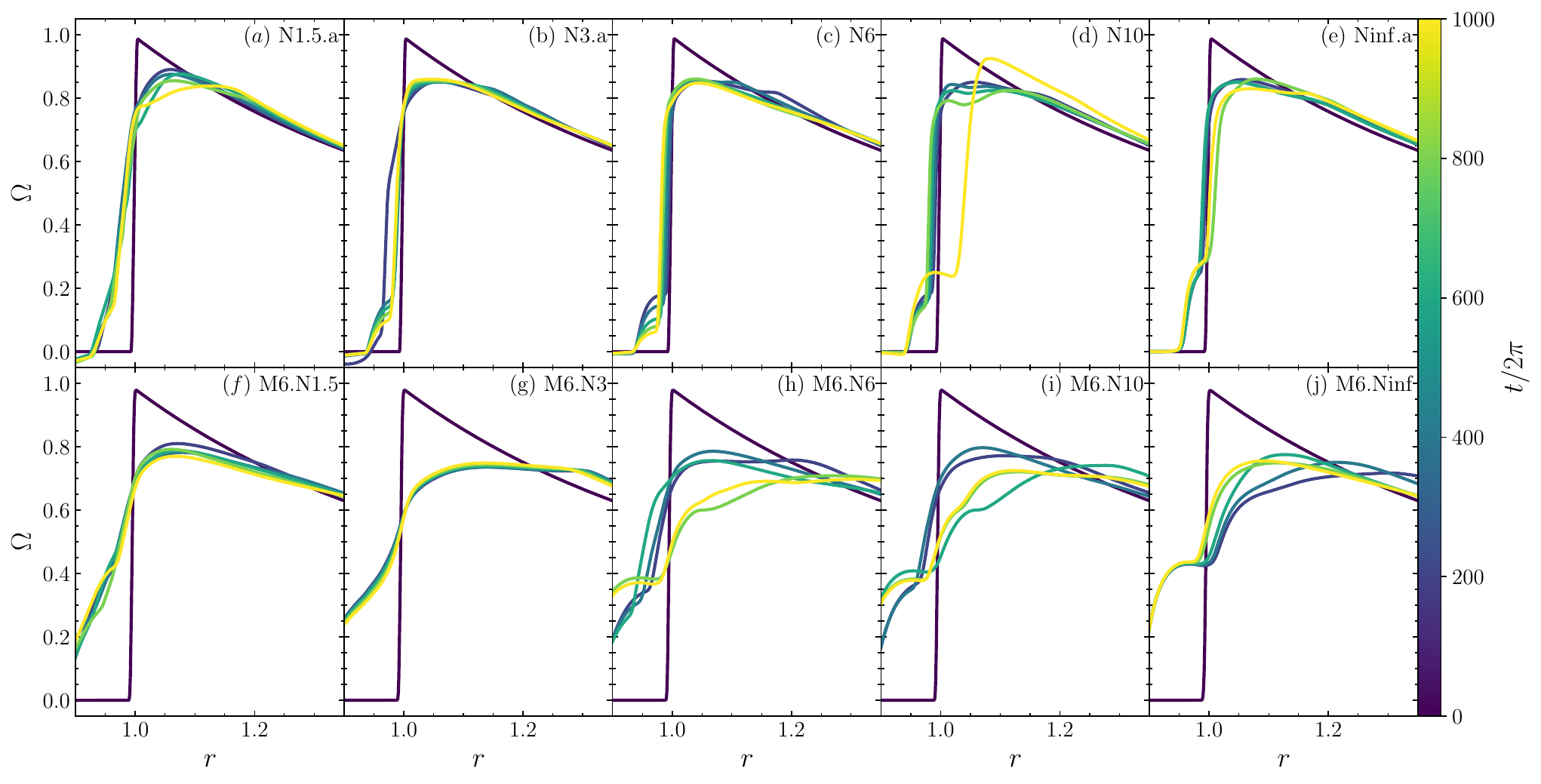}
    \caption{Profiles of ${\Omega(r)}$ for ten different simulations. The simulations chosen are for $n=1.5$, 3, 6, 10 and $\infty$ for $\mathcal{M}=6$ and 10. Full details of these simulations can be found in Table \ref{tab:all_sims}. In each panel, there are six profiles, taken at ${t/2\pi=0}$, 200, 400, 600, 800 and 1000.}
    \label{fig:Omega_profiles}
\end{figure*}

At this point, we can speculate that the dependence on stellar mass comes from the 'mode inertia' --- the volume integral of wave action (which has a density proportional to $\Sigma \mathrm{v}_r^2$) inside the star. This quantity is higher for deeper $r_\mathrm{in}$, which is also true for enclosed stellar mass. Looking at, for example, Figure \ref{fig:Ninf_snaps}, we can see that the acoustic modes are active throughout the star, right down to the inner boundary. This can be seen more clearly in Figure \ref{fig:depth}, which shows a snapshot of the wave action for three ${n=1.5}$ simulations with inner boundaries $r_\mathrm{in}$ placed at different depth. Each of these snapshots shows a mode which is active throughout the star, all the way to the respective $r_\mathrm{in}$. The higher mode inertia for deeper $r_\mathrm{in}$ may simply provide a larger reservoir of wave action that can couple to the modes in the disc, increasing the overall efficiency of acoustic mode excitation.

The slight discrepancy between the orange and grey points in the bottom panel of Figure \ref{fig:Mdot_trends_M} suggests that some other effects may also be at work here. Nevertheless, the correlation with the enclosed mass is still clearly present and strong and should be born in mind when interpreting these and other simulations.

\citet{Coleman+2022b} found that the magnitude of $\alpha_\mathrm{acc}$ and $\alpha_\mathrm{stress}$ does not depend on $\mathcal{M}$. To test this, Figure \ref{fig:Mdot_trends_M6} is analogous to Figure \ref{fig:Mdot_trends_M}, but shows the $\alpha$ values rather than $\dot{M}$ and $C_S$. The grey points refer to the same runs as in Figure \ref{fig:Mdot_trends_M}. Additionally, the red points show $\alpha$ values from the ${\mathcal{M}=6}$ simulations. In the top panel of \ref{fig:Mdot_trends_M6}, the red points lie on top of the trend of the grey and orange points, supporting the conclusion that $\alpha_\mathrm{acc}$ is independent of $\mathcal{M}$. However, the bottom panel clearly shows that the values of $\alpha_\mathrm{acc}$ from the ${\mathcal{M}=6}$ are greater in magnitude than those for ${\mathcal{M}=10}$ by a factor of a few. The reason for this is unclear, and would require simulations covering a greater range of Mach number to discern fully, which is beyond the scope of this work. %However, it is worth noting that the $\alpha$ values used in the \citet{Coleman+2022b} comparison were calculated at $r=1$, within the BL, whereas ours are calculated at $r=1.1$ within the inner disc.

%%%%%%%%%%%%%%%%%%%%%%%%%%%%%%%%%%%%
%%%%%%%%%%%%%%%%%%%%%%%%%%%%%%%%%%%%

\section{Radial Structure of the BL}
\label{sec:BL_structure}

%%%%%%%%%%%%%%%%%%%%%%%%%%%%%%%%%%%%

Finally, we will consider the sensitivity of the structure of the BL itself, specifically its angular velocity profile ${\Omega(r)}$, to the value of $n$ chosen to represent stellar structure. Figure \ref{fig:Omega_profiles} shows ${\Omega(r)}$ profiles at different times (spaced evenly by ${t/2\pi=200}$) for ten different simulations. The top row of panels corresponds to $\mathcal{M}=10$ and the bottom row $\mathcal{M}=6$. In each row, $n$ increases from left to right. The $\mathcal{M}=6$ simulations have much broader ${\Omega(r)}$ profiles, as found by \citet{Coleman+2022b}. In contrast, there is no obvious trend with $n$, something we investigate quantitatively shortly.

There is significant evolution from the initial condition (purple, $t=0$) to the next profile (slightly lighter blue, ${t/2\pi=200}$). However, beyond that point, there is often negligible evolution, which is especially apparent in Panels (a-c) and (e-g). While the other panels do show some later time evolution, it consists of $\Omega$ changes of order ${\sim10\%}$ in magnitude, and the qualitative description remains similar. It therefore appears that, once the initial shape of the $\Omega(r)$ profile is established (predominantly in an initial burst of accretion), the subsequent evolution of the system occurs with a relatively constant $\Omega(r)$.

All the simulations show super-Keplerian flow at radii ${r\gtrsim1.2}$. This arises due to the gap that has been created in the surface density profile. In this region ${\mathrm{d}\Sigma/\mathrm{d}r>0}$ and so there is an additional inwards force due to pressure gradient. A number of the ${\Omega(r)}$ profiles show humps in the inner BL. Perhaps the clearest example of this can be seen in Panel (d) (for the N10 simulation), where the yellow curve (at ${t/2\pi=1000}$) shows a large hump covering ${0.95\lesssim r\lesssim1.03}$. Where present, these humps penetrate into the stellar region, where the density is greater than within the disc. As a result, the angular momentum density in these regions is significant when compared to that in the disc, despite the modest values of $\Omega$. We therefore suggest that these humps arise as a result of the inward transport of mass and angular momentum from the disc, which must spin up the outer layers of the star. It is interesting that, for the ${\mathcal{M}=10}$ simulations (top row of Figure \ref{fig:Omega_profiles}), the humps never penetrate deeper than ${r=0.95}$. This value is strikingly similar to the final position of the passive scalar fields shown in Figures \ref{fig:passive_scalar} and \ref{fig:passive_scalar_int}. This supports the idea that the hump arises due to the accretion process. It further suggests that, while the accretion bursts discussed in Section \ref{sec:burst} can accrete material (and angular momentum) into the star, it can only do so into the outer layers, and not more deeply.

\begin{figure}
    \centering
    \includegraphics[width=\columnwidth]{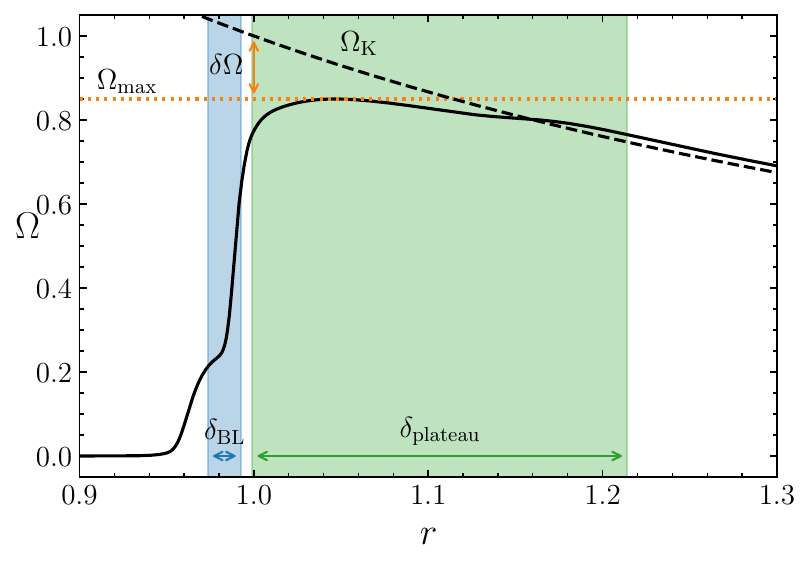}
    \caption{Angular velocity profile (solid black) taken at ${t/2\pi=1000}$ from the ${\mathcal{M}=10}$ globally isothermal simulation (Ninf.a), showing the definitions of $\delta_\mathrm{BL}$, $\delta\Omega$ and $\delta_\mathrm{plateau}$. Also shown are the Keplerian angular velocity (dashed black) and value of $\Omega_\mathrm{max}$ (dotted orange).}
    \label{fig:delta_defs}
\end{figure}

\citet{Coleman+2022b} found that the width of the BL was strongly dependent on the value of $\mathcal{M}$ in the simulations. This was understood by considering the local scale height within the star, which they concluded was responsible for setting the width of the BL. Looking at eq. \eqref{eq:star_den}, we can find that the relevant scale height at $r=1$ is given by ${(1+1/n)\mathcal{M}^{-2}}$. If this understanding is correct then, in contrast with the strong scaling of the BL width with $\mathcal{M}$, we should expect to see only weak scaling with $n$ as there is only a factor 2 difference in the scale height from ${n=1}$ to ${n=\infty}$. In order to test this scaling, we will define three measures of the shape of $\Omega(r)$ in the BL, which can then be compared across all our simulations. These measures are identical to those used in \citet{Coleman+2022a} and are illustrated in Figure \ref{fig:delta_defs}. 

First we have the width of the BL itself, defined as $\delta_\mathrm{BL} \equiv r\left(\Omega=0.75\Omega_\mathrm{max}\right) - r\left(\Omega=0.25\Omega_\mathrm{max}\right)$, where $\Omega_\mathrm{max}$ is simply the maximum value of $\Omega(r)$. Second we can quantify the suppression in the peak of $\Omega(r)$ by considering the difference between $\Omega_\mathrm{max}$ and the value of the Keplerian angular velocity ${\Omega_\mathrm{K}(r=1)=1}$. From this we have $\delta\Omega \equiv 1 - \Omega_\mathrm{max}$. Finally, we quantify the size of the plateau in $\Omega(r)$ by defining $\delta_\mathrm{plateau}$ as the radial extent for which ${\Omega(r)>0.9\Omega_\mathrm{max}}$. 

\begin{figure}
    \centering
    \includegraphics[width=\columnwidth]{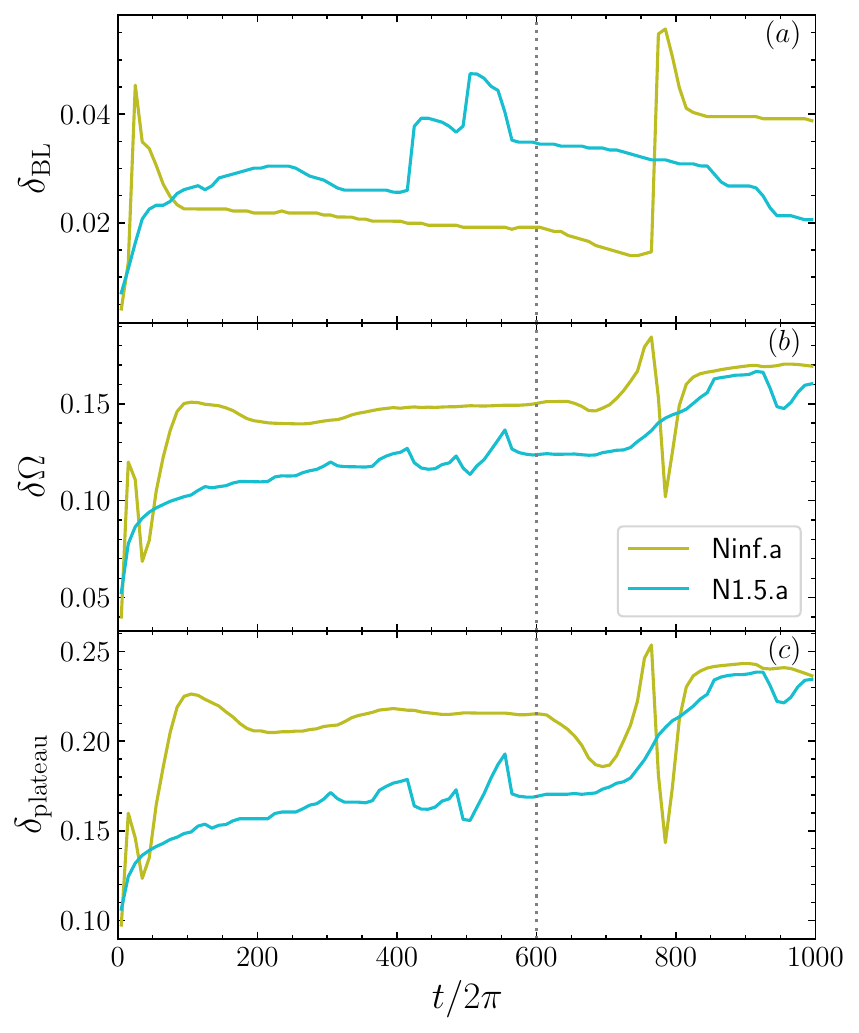}
    \caption{Temporal evolution of $\delta_\mathrm{BL}$ (a), $\delta\Omega$ (b) and $\delta_\mathrm{plateau}$ (c) (see text for definitions), for the simulations Ninf.a (green) and N1.5.a (blue). The vertical dotted lines at ${t/2\pi=600}$ correspond to the time at which the values used in Figure \ref{fig:delta_BL} are calculated.  }
    \label{fig:delta_BL_ev}
\end{figure}

Before we compare the values of these variables across all our simulations, it is useful to explore their temporal evolution.  Figure \ref{fig:delta_BL_ev} shows this evolution for the two simulations (Ninf.a and N1.5.a) discussed in Section \ref{sec:overview}. This reveals  temporal variation, by around a factor of 2, in all the quantities. Interestingly, the periods which show sudden variation (e.g. between ${t/2\pi=775}$ and 800 for Ninf.a) match directly with significant bursts of accretion (see Figures \ref{fig:Ninf_profiles} and \ref{fig:N1_5_profiles}). Beneath the stochastic events, there is no clear trend with time in either simulation, much as we saw in Figure \ref{fig:Omega_profiles}.

\begin{figure}
    \centering
    \includegraphics[width=\columnwidth]{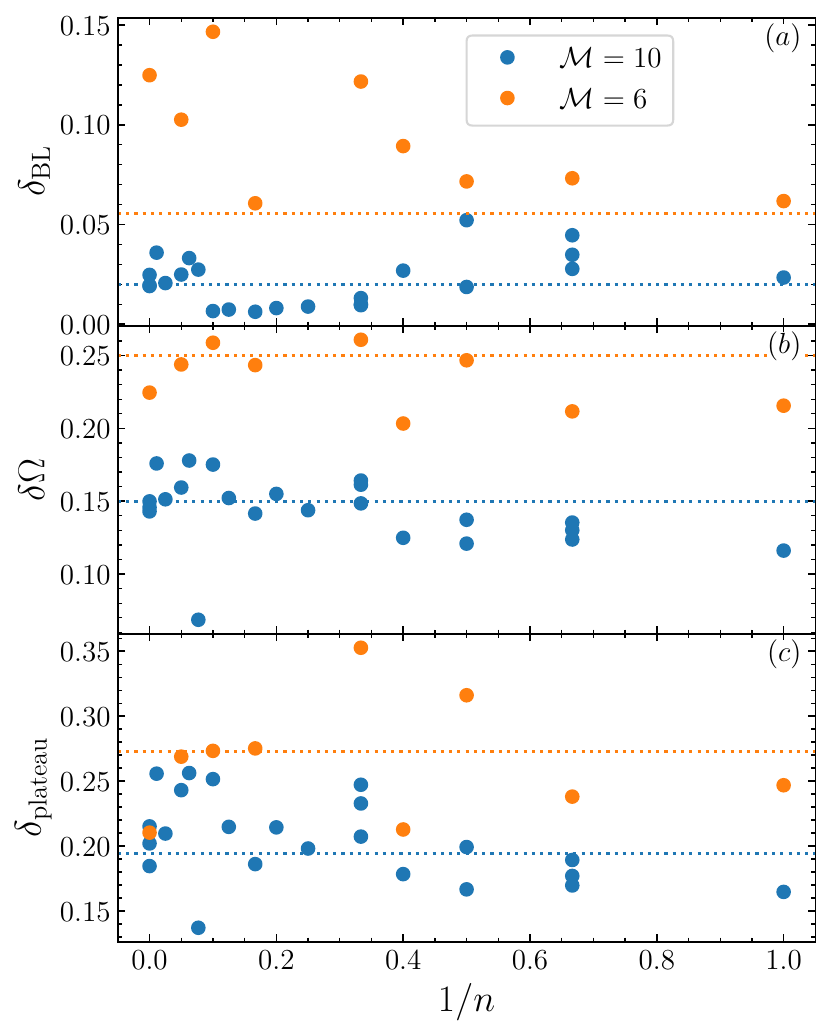}
    \caption{Plots of the BL characteristics $\delta_\mathrm{BL}$ (a), $\delta\Omega$ (b) and $\delta_\mathrm{plateau}$ (c). In each panel, the results from the ${\mathcal{M}=10}$ (blue) and 6 (orange) are shown. The dotted lines show the values  of the three quantities (for the respective Mach numbers) from the approximate trends against $\mathcal{M}$ found by \citet{Coleman+2022b} and given by ${\delta_\mathrm{BL}\approx2\mathcal{M}^{-2}}$, ${\delta\Omega\approx1.5\mathcal{M}^{-1}}$ and ${\delta_\mathrm{plateau}\approx0.9\mathcal{M}^{-2/3}}$. }
    \label{fig:delta_BL}
\end{figure}

We calculate the three quantities across our simulations at ${t/2\pi=600}$, the results of which are shown in Figure \ref{fig:delta_BL} as a function of $1/n$ for ${\mathcal{M}=10}$ (blue) and 6 (orange). As argued above, this choice of time does not make a significant difference to the trends seen, but this was the time chosen by \citet{Coleman+2022b}\footnote{This choice corresponded to the end of the simulations presented in \citet{Coleman+2022a, Coleman+2022b}.} and so this choice allows for a direct comparison between our work and theirs. Indeed, the horizontal dotted lines in Figure \ref{fig:delta_BL} show the values found by \citet{Coleman+2022b} for globally isothermal simulations, details of which are in the figure caption.

Looking first at our globally isothermal simulations, for which ${1/n=0}$, there is good agreement between our values and those from \citet{Coleman+2022b}, with the exception of the value of $\delta_\mathrm{BL}$ for the ${\mathcal{M}=6}$ simulation, for which our value is more than twice that from \citet{Coleman+2022b}. Looking more generally across all simulations, there is no clear trend in any of the three variables with $n$. This is perhaps unsurprising given that, as mentioned earlier, the scale height within the star is given by ${(1+1/n)\mathcal{M}^{-2}}$. Any scaling with $n$ would therefore be much weaker than with $\mathcal{M}$. Whether there is a trend with $n$ or not is not possible to ascertain from this data, given its stochastic nature, but it is clear that any trend would be weak.

%%%%%%%%%%%%%%%%%%%%%%%%%%%%%%%%%%%%
%%%%%%%%%%%%%%%%%%%%%%%%%%%%%%%%%%%%

\section{Discussion} \label{sec:discussion}

%%%%%%%%%%%%%%%%%%%%%%%%%%%%%%%%%%%%

In this work, we have presented a suite of 2D simulations to explore the behaviour and evolution of BL accretors. We have focussed on the effect that the stellar structure, expressed through the polytropic index $n$, has on the acoustic mode driven accretion. To the best of our knowledge, this represents the first time that the underlying stellar temperature profile of BL simulations has not been globally isothermal. As such, our study is a key step in moving towards more realistic models of BL systems.

We have shown that varying the polytropic index $n$ within the star has a number of effects on the emergent phenomena. As noted earlier, the scale height of the stellar density near ${r=1}$ is given by ${(1+1/n)\mathcal{M}^{-2}}$, which is a relatively weak function of $n$.  In contrast, the global stellar structure, quantified by the total mass within the simulation domain, depends heavily on $n$. This can be seen, for example, by considering the radial range of Figure \ref{fig:Mdot_trends_M}, where the difference between ${n=1.5}$ and ${n\rightarrow\infty}$ is five orders of magnitude.

A priori, we might therefore expect that variables which depend primarily on the local structure of the BL are only weakly dependent on $n$, whereas those that depend on the global structure of the star vary significantly with $n$. In Section \ref{sec:BL_structure}, we showed that the $\Omega(r)$ profile across the BL showed no significant variation with $n$. This is consistent with the conclusions of \citet{Coleman+2022b}, who showed that the BL structure did vary with $\mathcal{M}$ in a manner consistent with it being set by the local scale height.

The mix of acoustic modes excited by the BL might also be expected to depend on the local structure. In the original analytical work \citep{Belyaev&Rafikov2012}, there is no requirement for a density contrast, or any dependence on the density profile through the shear layer. Nevertheless, \citet{Coleman+2022a} found that the mix of modes was a function of the Mach number of the flow, and thus we might expect something similar with $n$. In  Section \ref{sec:mode_mix}, we explored the acoustic modes present. We found that the average azimuthal wavenumber ${\langle m\rangle}$ of modes present in the disc does increase with $n$. However, no such similar relation was found for the stellar modes. The trend of ${\langle m\rangle}$ with $n$ is relatively weak, and suggests that it does vary with the local scale height (which also varies weakly with $n$) rather than the global stellar structure (which varies strongly).

On the other hand, in Section \ref{sec:accretion} we found the near-BL mass and angular momentum transport to be a strong function of $n$. Additionally, varying the location of the inner boundary whilst keeping $n$ constant also affected the magnitude of the transport. This, alongside the findings of \citet{Coleman+2022b} that this transport is independent of $\mathcal{M}$, suggest that it depends on the global structure of the star. The exact mechanism by which this occurs is not clear, but it is certainly true that the acoustic modes within the star penetrate all the way to the inner boundary, even in simulations with the deepest $r_\mathrm{in}$, see Figure \ref{fig:depth}. This suggests that the modes have a way of sampling the full stellar structure, which influences their evolution. However, complicating this picture is the fact that, as seen in Figure \ref{fig:all_modes}, the vast majority of modes detected in our simulations are seen only in the star or the disc (i.e. they appear to be not fully global). This may be caused by the different amplitudes of a global mode in the two regions or the ambiguity in its detection across the full domain. One should also bear in mind that our simulations are not designed to reach a constant $\dot M$ steady state, which may affect the interpretation of their results. Whatever the underlying reason behind the trend of transport processes with the enclosed stellar mass, it is robustly observed in this work and so worthy of further study.

%%%%%%%%%%%%%%%%%%%%%%%%%%%%%%%%%%%%

\subsection{Comparison to 3D} 
\label{sec:3D}

%%%%%%%%%%%%%%%%%%%%%%%%%%%%%%%%%%%%

\begin{figure}
    \centering
    \includegraphics[width=\columnwidth]{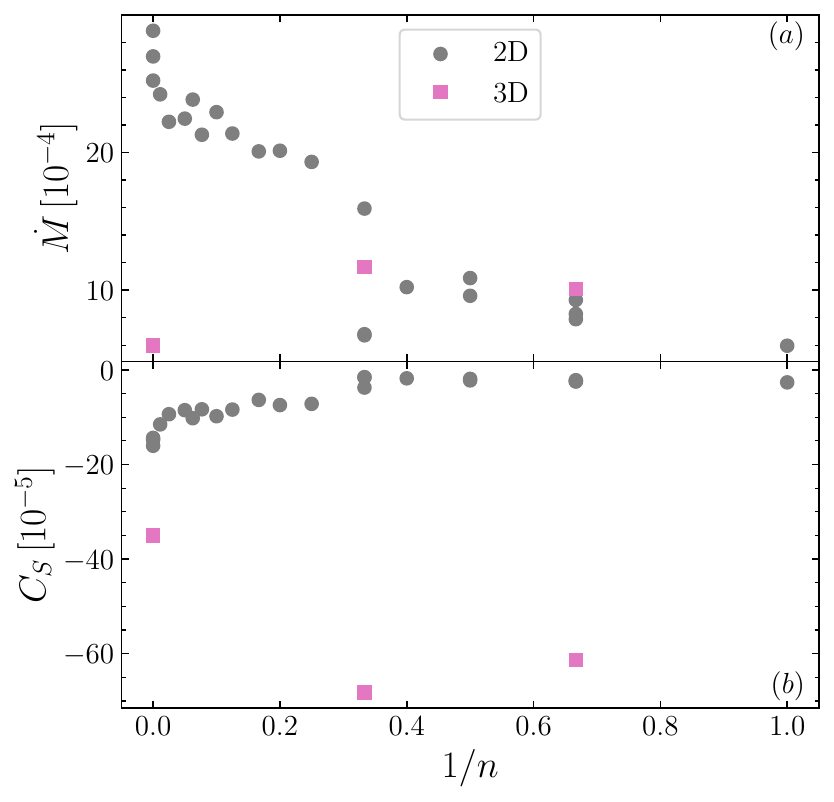}
    \caption{As for Figure \ref{fig:Mdot_trends_ninv}, showing the accretion rate (a) and wave driven angular momentum transport (b) as a function of $1/n$, calculated for the first ${t/2\pi=100}$. The grey points are from 2D simulations as in Figure \ref{fig:Mdot_trends_ninv}, and the pink squares show values calculated from 3D simulations, where the vertical integration has been performed over 2 scale heights.  }
    \label{fig:Mdot_trends_3D}
\end{figure}

Everything discussed thus far has been based on our suite of 2D simulations. It does, however, go without saying that real accreting systems are not well described by the 2D approximation. The advantage of restricting this study to 2D lies in the relatively low computational cost, which allows the thorough exploration of parameter space that we have performed. Nevertheless, it is useful to consider potential differences with 3D simulations. To do this, we briefly compare the mass and angular momentum transport for our ${\mathcal{M}=10}$ simulations (as presented in Figure \ref{fig:Mdot_trends_ninv}, with preliminary results from three 3D ${\mathcal{M}=10}$ simulations, with $n=1.5$, 3 and $\infty$. These simulations were run in full spherical geometry (fully covering the outer layers of the star) with \texttt{Athena++} for a total duration of ${t/2\pi=500}$; a thorough analysis of their evolution and properties will be presented in a future work. 

From these simulations, we calculate average values of $\dot{M}$ and $C_S$ at ${r=1.1}$, again averaging over the initial ${t/2\pi=100}$ as for the 2D results. The values were calculated by vertically integrating over two scale heights within the disc. The results are shown in Figure \ref{fig:Mdot_trends_3D}, along with the grey points, which are the same as the grey points in Figure \ref{fig:Mdot_trends_ninv}. We choose to plot against ${1/n}$, rather than what we argued is the more fundamental ${M(r\leq1)}$, due to the complexity is determining ${M(r\leq1)}$ for the 3D simulations. The difference in the vertical structure between the star and the disc makes the process non-trivial. For this reason, we also omit the orange points from Figure \ref{fig:Mdot_trends_ninv}, which did not fit the trend against ${1/n}$ (but did with ${M(r\leq1)}$). Despite this, the three 3D simulations have the same inner boundary as the 2D simulations and so a direct comparison against the ${1/n}$ domain is valid.

It is immediately obvious that the 3D results tell a different story to 2D. Starting with the accretion rate, the $n=1.5$ and 3 simulations are in good agreement with the 2D trend, but the globally isothermal simulation (${1/n=0}$) is not. The initial ${t/2\pi=100}$ of the globally isothermal simulation was rerun with a different random seed, to see if this discrepancy is due to stochasticity. Although not shown on Figure \ref{fig:Mdot_trends_3D}, the values for this rerun are almost identical to the original, suggesting that the discrepancy is real. The discrepancy between the $C_S$ values is even greater, differing by up to an order of magnitude. Unlike the 2D case, there is no obvious trend with ${1/n}$ in either variable, although this results is limited by only having three simulations.

The reason behind this discrepancy is not clear, although we caution against reading too much into these preliminary results. The values of $\dot{M}$ and $C_S$ vary significantly depending on the range over which the vertical integration is performed. This suggests that the full 3D nature of the flow is critically important for understanding the behaviour, given that phenomena such as meridional circulation, impossible in 2D simulations, can be present in 3D. We therefore leave a thorough analysis and explanation of this to future work.

%%%%%%%%%%%%%%%%%%%%%%%%%%%%%%%%%%%%
 
\subsection{Comparison to Previous Work}

%%%%%%%%%%%%%%%%%%%%%%%%%%%%%%%%%%%%

This paper is by no means the first to explore accretion through the BL. A large number of past studies relied on a local viscosity within the BL to drive accretion \citep{Kippenhahn&Thomas1978, Narayan&Popham1993, Popham+1993, Kley&Lin1996, Piro&Bildsten2004, Balsara+2009, Hertfelder+2013, Dong+2021}. The intrinsically global nature of the acoustic modes \citep{Belyaev+2012, Belyaev+2013a} means that we cannot make direct comparisons to these studies.

This work builds on previous HD simulations of BL systems \citep{Belyaev+2012, Belyaev+2013a,Coleman+2022a, Coleman+2022b}, which explored the details of acoustic mode-driven accretion. These studies have been extended in many ways since then, by including a non-zero angular velocity of the star \citep{Dittmann2021}, non-trivial thermodynamics \citep{Hertfelder&Kley2015, Dittmann2021}, effects of MHD and MRI in 3D \citep{Belyaev+2013b, Belyaev&Quataert2018,Takasao2025}. Our work is complementary to these studies in that it explores a different dimension of the problem --- the effect of the stellar structure on the mode operation. 

%%%%%%%%%%%%%%%%%%%%%%%%%%%%%%%%%%%%

\subsection{Caveats}

%%%%%%%%%%%%%%%%%%%%%%%%%%%%%%%%%%%%

In order to make our study tractable, we have made a number of assumptions and simplifications, which we detail briefly. Firstly, our simulations have been restricted to 2D. This simplification is required to perform the sweep across the parameter space. However, an initial comparison to data from 3D simulations (see Section \ref{sec:3D}) suggests that the picture is more complex when the vertical structure of the system is considered.

We have additionally neglected any influence from magnetic fields on the system. Previous work examining MHD simulations has found that the acoustic modes are still present with magnetic fields \citep{Belyaev+2013b, Belyaev&Quataert2018}. Perhaps the biggest limitation to not including magnetic fields is that the inner disc is not refilled. In real systems, angular momentum transport in the disc is expected to be provided by the MRI. This would provided a continuous supply of material to the inner regions. In our simulations, this does not occur. Instead, once material has been accreted from the inner disc and a gap in the ${\Sigma(r)}$ profile created, there is no mechanism by which it can be replenished. It is therefore likely that the long-term evolution of our simulations is missing this important process. One way we have attempted to mitigate this is by considering the mass accretion and angular momentum transport over the initial ${t/2\pi=100}$ of our simulations. In this period, the inner disc is still relatively full of material, and thus may be more representative of a true system with steady mass replenishment.

The final caveat which is important to discuss is the thermodynamics within the simulation. In this work, we use a locally isothermal EoS for treating the evolution of perturbations, with the background temperature that is generally not spatially constant. While this is certainly more realistic than a uniform background temperature used in a number of earlier studies, it is still a gross approximation. Radiative cooling is known to have a significant effect on density wave evolution in discs \citep{Miranda2020a,Miranda2020b, Zhang2020, Ziampras2023}, with efficient damping of the wave amplitude at intermediate cooling timescales. \citet{Dittmann2024} explored the effect of a non-isothermal EoS with a $\beta$-cooling model for treating the perturbations and found it to play an important role. 

In real systems, the disc temperature may also far exceed the surface temperature of the star. A realistic temperature jump across the BL would also require a jump in the density, since the pressure must be continuous across the BL. Therefore, real systems would likely have a much larger star-to-disc mass ratio than used here. This could be very important, given that we showed in Section \ref{sec:accretion} that the stellar mass is crucial in determining the accretion and angular momentum transport. The inclusion of more realistic thermodynamics, including a temperature jump across the BL, is left to future work.

%%%%%%%%%%%%%%%%%%%%%%%%%%%%%%%%%%%%
%%%%%%%%%%%%%%%%%%%%%%%%%%%%%%%%%%%%

\section{Conclusions} 
\label{sec:conclusions}

%%%%%%%%%%%%%%%%%%%%%%%%%%%%%%%%%%%%

In this work, we explored the effect that the inner temperature and density structure of an accreting object with a surface has on the operation of the acoustic modes excited in the boundary layer. By parametrizing the stellar structure as a polytropic sphere with an index $n$, we explored a range of stellar density and temperature profiles, from steeply rising density for $n=\infty$ to much slower increasing density at lower values of $n$. We find that acoustic modes get excited in the BL regardless of the deep stellar structure (the value on $n$), with roughly the same mix of modes as in other studies, although we do find more high-$m$ lower modes for lower $n$. However, the efficiencies of mass and angular momentum transport, measured through the mass accretion rate and the angular momentum flux across the BL, are both found to be strong functions of $n$. Interestingly, in 2D the tightest correlation of the transport metrics is not with $n$ but with the total mass inside the star that lies within the simulation domain. We suggest that this correlation has a physical (rather than a numerical) origin, perhaps related to the 'mode inertia' inside the star playing an important role is exciting the acoustic modes in the BL (see Section \ref{sec:accretion}). We also discuss the evolution of the near-BL region during an intense accretion burst and briefly show some preliminary results of the 3D simulations of the BLs, which will be described in more details in future work.

%%%%%%%%%%%%%%%%%%%%%%%%%%%%%%%%%%%%
%%%%%%%%%%%%%%%%%%%%%%%%%%%%%%%%%%%%
 
\section*{Acknowledgements}

%%%%%%%%%%%%%%%%%%%%%%%%%%%%%%%%%%%%

We are grateful to the referee, Matthew Coleman, for their careful review and numerous suggestions that helped improve this manuscript. We are also indebted to Callum Fairbairn for their help with Python. Authors acknowledge financial support through the STFC grant ST/T00049X/1; RRR was also supported by the IAS.
This work used the DiRAC Memory Intensive service (Cosma8) at Durham University, managed by the Institute for Computational Cosmology on behalf of the STFC DiRAC HPC Facility (www.dirac.ac.uk). The DiRAC service at Durham was funded by BEIS, UKRI and STFC capital funding, Durham University and STFC operations grants. DiRAC is part of the UKRI Digital Research Infrastructure.\\
\textit{Software:} NumPy \citep{harris_array_2020}, SciPy \citep{virtanen_scipy_2020}, Matplotlib \citep{thomas_a_caswell_matplotlibmatplotlib_2023} and Athena++ \citep{Stone+2020}.

%%%%%%%%%%%%%%%%%%%%%%%%%%%%%%%%%%%%
%%%%%%%%%%%%%%%%%%%%%%%%%%%%%%%%%%%%

\section*{Data Availability}

%%%%%%%%%%%%%%%%%%%%%%%%%%%%%%%%%%%%

Simulation data will be made available upon reasonable request to the corresponding author.



\bibliographystyle{mnras}
\bibliography{references} % if your bibtex file is called example.bib

\begin{thebibliography}{}
\makeatletter
\relax
\def\mn@urlcharsother{\let\do\@makeother \do\$\do\&\do\#\do\^\do\_\do\%\do\~}
\def\mn@doi{\begingroup\mn@urlcharsother \@ifnextchar [ {\mn@doi@}
  {\mn@doi@[]}}
\def\mn@doi@[#1]#2{\def\@tempa{#1}\ifx\@tempa\@empty \href
  {http://dx.doi.org/#2} {doi:#2}\else \href {http://dx.doi.org/#2} {#1}\fi
  \endgroup}
\def\mn@eprint#1#2{\mn@eprint@#1:#2::\@nil}
\def\mn@eprint@arXiv#1{\href {http://arxiv.org/abs/#1} {{\tt arXiv:#1}}}
\def\mn@eprint@dblp#1{\href {http://dblp.uni-trier.de/rec/bibtex/#1.xml}
  {dblp:#1}}
\def\mn@eprint@#1:#2:#3:#4\@nil{\def\@tempa {#1}\def\@tempb {#2}\def\@tempc
  {#3}\ifx \@tempc \@empty \let \@tempc \@tempb \let \@tempb \@tempa \fi \ifx
  \@tempb \@empty \def\@tempb {arXiv}\fi \@ifundefined
  {mn@eprint@\@tempb}{\@tempb:\@tempc}{\expandafter \expandafter \csname
  mn@eprint@\@tempb\endcsname \expandafter{\@tempc}}}

\bibitem[\protect\citeauthoryear{{Balbus} \& {Hawley}}{{Balbus} \&
  {Hawley}}{1991}]{Balbus&Hawley1991}
{Balbus} S.~A.,  {Hawley} J.~F.,  1991, \mn@doi [\apj] {10.1086/170270}, \href
  {https://ui.adsabs.harvard.edu/abs/1991ApJ...376..214B} {376, 214}

\bibitem[\protect\citeauthoryear{{Balbus} \& {Papaloizou}}{{Balbus} \&
  {Papaloizou}}{1999}]{Balbus&Papaloizou1999}
{Balbus} S.~A.,  {Papaloizou} J. C.~B.,  1999, \mn@doi [\apj] {10.1086/307594},
  \href {https://ui.adsabs.harvard.edu/abs/1999ApJ...521..650B} {521, 650}

\bibitem[\protect\citeauthoryear{{Balsara}, {Fisker}, {Godon}  \&
  {Sion}}{{Balsara} et~al.}{2009}]{Balsara+2009}
{Balsara} D.~S.,  {Fisker} J.~L.,  {Godon} P.,   {Sion} E.~M.,  2009, \mn@doi
  [\apj] {10.1088/0004-637X/702/2/1536}, \href
  {https://ui.adsabs.harvard.edu/abs/2009ApJ...702.1536B} {702, 1536}

\bibitem[\protect\citeauthoryear{{Belyaev}}{{Belyaev}}{2017}]{Belyaev2017}
{Belyaev} M.~A.,  2017, \mn@doi [\apj] {10.3847/1538-4357/835/2/238}, \href
  {https://ui.adsabs.harvard.edu/abs/2017ApJ...835..238B} {835, 238}

\bibitem[\protect\citeauthoryear{{Belyaev} \& {Quataert}}{{Belyaev} \&
  {Quataert}}{2018}]{Belyaev&Quataert2018}
{Belyaev} M.~A.,  {Quataert} E.,  2018, \mn@doi [\mnras]
  {10.1093/mnras/sty860}, \href
  {https://ui.adsabs.harvard.edu/abs/2018MNRAS.479.1528B} {479, 1528}

\bibitem[\protect\citeauthoryear{{Belyaev} \& {Rafikov}}{{Belyaev} \&
  {Rafikov}}{2012}]{Belyaev&Rafikov2012}
{Belyaev} M.~A.,  {Rafikov} R.~R.,  2012, \mn@doi [\apj]
  {10.1088/0004-637X/752/2/115}, \href
  {https://ui.adsabs.harvard.edu/abs/2012ApJ...752..115B} {752, 115}

\bibitem[\protect\citeauthoryear{{Belyaev}, {Rafikov}  \& {Stone}}{{Belyaev}
  et~al.}{2012}]{Belyaev+2012}
{Belyaev} M.~A.,  {Rafikov} R.~R.,   {Stone} J.~M.,  2012, \mn@doi [\apj]
  {10.1088/0004-637X/760/1/22}, \href
  {https://ui.adsabs.harvard.edu/abs/2012ApJ...760...22B} {760, 22}

\bibitem[\protect\citeauthoryear{{Belyaev}, {Rafikov}  \& {Stone}}{{Belyaev}
  et~al.}{2013a}]{Belyaev+2013a}
{Belyaev} M.~A.,  {Rafikov} R.~R.,   {Stone} J.~M.,  2013a, \mn@doi [\apj]
  {10.1088/0004-637X/770/1/67}, \href
  {https://ui.adsabs.harvard.edu/abs/2013ApJ...770...67B} {770, 67}

\bibitem[\protect\citeauthoryear{{Belyaev}, {Rafikov}  \& {Stone}}{{Belyaev}
  et~al.}{2013b}]{Belyaev+2013b}
{Belyaev} M.~A.,  {Rafikov} R.~R.,   {Stone} J.~M.,  2013b, \mn@doi [\apj]
  {10.1088/0004-637X/770/1/68}, \href
  {https://ui.adsabs.harvard.edu/abs/2013ApJ...770...68B} {770, 68}

\bibitem[\protect\citeauthoryear{Caswell et~al.,}{Caswell
  et~al.}{2023}]{thomas_a_caswell_matplotlibmatplotlib_2023}
Caswell T.~A.,  et~al., 2023, matplotlib/matplotlib: {REL}: v3.7.4,
  \mn@doi{10.5281/ZENODO.592536}, \url
  {https://zenodo.org/doi/10.5281/zenodo.592536}

\bibitem[\protect\citeauthoryear{{Coleman}, {Rafikov}  \&
  {Philippov}}{{Coleman} et~al.}{2022a}]{Coleman+2022a}
{Coleman} M. S.~B.,  {Rafikov} R.~R.,   {Philippov} A.~A.,  2022a, \mn@doi
  [\mnras] {10.1093/mnras/stab2962}, \href
  {https://ui.adsabs.harvard.edu/abs/2022MNRAS.509..440C} {509, 440}

\bibitem[\protect\citeauthoryear{{Coleman}, {Rafikov}  \&
  {Philippov}}{{Coleman} et~al.}{2022b}]{Coleman+2022b}
{Coleman} M. S.~B.,  {Rafikov} R.~R.,   {Philippov} A.~A.,  2022b, \mn@doi
  [\mnras] {10.1093/mnras/stac732}, \href
  {https://ui.adsabs.harvard.edu/abs/2022MNRAS.512.2945C} {512, 2945}

\bibitem[\protect\citeauthoryear{{Cordwell} \& {Rafikov}}{{Cordwell} \&
  {Rafikov}}{2024}]{Cordwell2024}
{Cordwell} A.~J.,  {Rafikov} R.~R.,  2024, \mn@doi [\mnras]
  {10.1093/mnras/stae2089}, \href
  {https://ui.adsabs.harvard.edu/abs/2024MNRAS.534.1394C} {534, 1394}

\bibitem[\protect\citeauthoryear{{Dittmann}}{{Dittmann}}{2021}]{Dittmann2021}
{Dittmann} A.~J.,  2021, \mn@doi [\mnras] {10.1093/mnras/stab2682}, \href
  {https://ui.adsabs.harvard.edu/abs/2021MNRAS.508.1842D} {508, 1842}

\bibitem[\protect\citeauthoryear{{Dittmann}}{{Dittmann}}{2024}]{Dittmann2024}
{Dittmann} A.~J.,  2024, \mn@doi [\apj] {10.3847/1538-4357/ad6d6b}, \href
  {https://ui.adsabs.harvard.edu/abs/2024ApJ...974..218D} {974, 218}

\bibitem[\protect\citeauthoryear{{Dong}, {Jiang}  \& {Armitage}}{{Dong}
  et~al.}{2021}]{Dong+2021}
{Dong} J.,  {Jiang} Y.-F.,   {Armitage} P.~J.,  2021, \mn@doi [\apj]
  {10.3847/1538-4357/ac1941}, \href
  {https://ui.adsabs.harvard.edu/abs/2021ApJ...921...54D} {921, 54}

\bibitem[\protect\citeauthoryear{{Drury}}{{Drury}}{1980}]{Drury1980}
{Drury} L.~O.,  1980, \mn@doi [\mnras] {10.1093/mnras/193.2.337}, \href
  {https://ui.adsabs.harvard.edu/abs/1980MNRAS.193..337D} {193, 337}

\bibitem[\protect\citeauthoryear{{Drury}}{{Drury}}{1985}]{Drury1985}
{Drury} L.~O.,  1985, \mn@doi [\mnras] {10.1093/mnras/217.4.821}, \href
  {https://ui.adsabs.harvard.edu/abs/1985MNRAS.217..821D} {217, 821}

\bibitem[\protect\citeauthoryear{{Ferland}, {Langer}, {MacDonald}, {Pepper},
  {Shaviv}  \& {Truran}}{{Ferland} et~al.}{1982}]{Ferland+1982}
{Ferland} G.~J.,  {Langer} S.~H.,  {MacDonald} J.,  {Pepper} G.~H.,  {Shaviv}
  G.,   {Truran} J.~W.,  1982, \mn@doi [\apjl] {10.1086/183910}, \href
  {https://ui.adsabs.harvard.edu/abs/1982ApJ...262L..53F} {262, L53}

\bibitem[\protect\citeauthoryear{{Fromang} \& {Nelson}}{{Fromang} \&
  {Nelson}}{2006}]{Fromang&Nelson2006}
{Fromang} S.,  {Nelson} R.~P.,  2006, \mn@doi [\aap]
  {10.1051/0004-6361:20065643}, \href
  {https://ui.adsabs.harvard.edu/abs/2006A&A...457..343F} {457, 343}

\bibitem[\protect\citeauthoryear{{Fu}, {Huang}  \& {Yu}}{{Fu}
  et~al.}{2023}]{Fu+2023}
{Fu} Z.,  {Huang} S.,   {Yu} C.,  2023, \mn@doi [\apj]
  {10.3847/1538-4357/acac9c}, \href
  {https://ui.adsabs.harvard.edu/abs/2023ApJ...945..165F} {945, 165}

\bibitem[\protect\citeauthoryear{{Ghosh} \& {Lamb}}{{Ghosh} \&
  {Lamb}}{1978}]{Ghosh&Lamb1978}
{Ghosh} P.,  {Lamb} F.~K.,  1978, \mn@doi [\apjl] {10.1086/182734}, \href
  {https://ui.adsabs.harvard.edu/abs/1978ApJ...223L..83G} {223, L83}

\bibitem[\protect\citeauthoryear{{Gilfanov}, {Revnivtsev}  \&
  {Molkov}}{{Gilfanov} et~al.}{2003}]{Gilfanov+2003}
{Gilfanov} M.,  {Revnivtsev} M.,   {Molkov} S.,  2003, \mn@doi [\aap]
  {10.1051/0004-6361:20031141}, \href
  {https://ui.adsabs.harvard.edu/abs/2003A&A...410..217G} {410, 217}

\bibitem[\protect\citeauthoryear{{Glatzel}}{{Glatzel}}{1988}]{Glatzel1988}
{Glatzel} W.,  1988, \mn@doi [\mnras] {10.1093/mnras/231.3.795}, \href
  {https://ui.adsabs.harvard.edu/abs/1988MNRAS.231..795G} {231, 795}

\bibitem[\protect\citeauthoryear{{Goldreich} \& {Tremaine}}{{Goldreich} \&
  {Tremaine}}{1978}]{Goldreich&Tremaine1978}
{Goldreich} P.,  {Tremaine} S.,  1978, \mn@doi [\apj] {10.1086/156203}, \href
  {https://ui.adsabs.harvard.edu/abs/1978ApJ...222..850G} {222, 850}

\bibitem[\protect\citeauthoryear{Harris et~al.,}{Harris
  et~al.}{2020}]{harris_array_2020}
Harris C.~R.,  et~al., 2020, \mn@doi [Nature] {10.1038/s41586-020-2649-2}, 585,
  357

\bibitem[\protect\citeauthoryear{{Hertfelder} \& {Kley}}{{Hertfelder} \&
  {Kley}}{2015}]{Hertfelder&Kley2015}
{Hertfelder} M.,  {Kley} W.,  2015, \mn@doi [\aap]
  {10.1051/0004-6361/201526005}, \href
  {https://ui.adsabs.harvard.edu/abs/2015A&A...579A..54H} {579, A54}

\bibitem[\protect\citeauthoryear{{Hertfelder}, {Kley}, {Suleimanov}  \&
  {Werner}}{{Hertfelder} et~al.}{2013}]{Hertfelder+2013}
{Hertfelder} M.,  {Kley} W.,  {Suleimanov} V.,   {Werner} K.,  2013, \mn@doi
  [\aap] {10.1051/0004-6361/201322542}, \href
  {https://ui.adsabs.harvard.edu/abs/2013A&A...560A..56H} {560, A56}

\bibitem[\protect\citeauthoryear{{Inogamov} \& {Sunyaev}}{{Inogamov} \&
  {Sunyaev}}{1999}]{Inogamov&Sunyaev1999}
{Inogamov} N.~A.,  {Sunyaev} R.~A.,  1999, \mn@doi [Astronomy Letters]
  {10.48550/arXiv.astro-ph/9904333}, \href
  {https://ui.adsabs.harvard.edu/abs/1999AstL...25..269I} {25, 269}

\bibitem[\protect\citeauthoryear{{Inogamov} \& {Sunyaev}}{{Inogamov} \&
  {Sunyaev}}{2010}]{Inogamov&Sunyaev2010}
{Inogamov} N.~A.,  {Sunyaev} R.~A.,  2010, \mn@doi [Astronomy Letters]
  {10.1134/S1063773710120029}, \href
  {https://ui.adsabs.harvard.edu/abs/2010AstL...36..848I} {36, 848}

\bibitem[\protect\citeauthoryear{{Johnson} \& {Gammie}}{{Johnson} \&
  {Gammie}}{2005}]{Johnson&Gammie2005}
{Johnson} B.~M.,  {Gammie} C.~F.,  2005, \mn@doi [\apj] {10.1086/497358}, \href
  {https://ui.adsabs.harvard.edu/abs/2005ApJ...635..149J} {635, 149}

\bibitem[\protect\citeauthoryear{{Kenyon}, {Hartmann}, {Imhoff}  \&
  {Cassatella}}{{Kenyon} et~al.}{1989}]{Kenyon+1989}
{Kenyon} S.~J.,  {Hartmann} L.,  {Imhoff} C.~L.,   {Cassatella} A.,  1989,
  \mn@doi [\apj] {10.1086/167860}, \href
  {https://ui.adsabs.harvard.edu/abs/1989ApJ...344..925K} {344, 925}

\bibitem[\protect\citeauthoryear{{Kippenhahn} \& {Thomas}}{{Kippenhahn} \&
  {Thomas}}{1978}]{Kippenhahn&Thomas1978}
{Kippenhahn} R.,  {Thomas} H.~C.,  1978, \aap, \href
  {https://ui.adsabs.harvard.edu/abs/1978A&A....63..265K} {63, 265}

\bibitem[\protect\citeauthoryear{{Kley} \& {Lin}}{{Kley} \&
  {Lin}}{1996}]{Kley&Lin1996}
{Kley} W.,  {Lin} D.~N.~C.,  1996, \mn@doi [\apj] {10.1086/177115}, \href
  {https://ui.adsabs.harvard.edu/abs/1996ApJ...461..933K} {461, 933}

\bibitem[\protect\citeauthoryear{{Li}, {Finn}, {Lovelace}  \& {Colgate}}{{Li}
  et~al.}{2000}]{Li+2000}
{Li} H.,  {Finn} J.~M.,  {Lovelace} R.~V.~E.,   {Colgate} S.~A.,  2000, \mn@doi
  [\apj] {10.1086/308693}, \href
  {https://ui.adsabs.harvard.edu/abs/2000ApJ...533.1023L} {533, 1023}

\bibitem[\protect\citeauthoryear{{Li}, {Colgate}, {Wendroff}  \& {Liska}}{{Li}
  et~al.}{2001}]{Li+2001}
{Li} H.,  {Colgate} S.~A.,  {Wendroff} B.,   {Liska} R.,  2001, \mn@doi [\apj]
  {10.1086/320241}, \href
  {https://ui.adsabs.harvard.edu/abs/2001ApJ...551..874L} {551, 874}

\bibitem[\protect\citeauthoryear{{Lovelace}, {Li}, {Colgate}  \&
  {Nelson}}{{Lovelace} et~al.}{1999}]{Lovelace+1999}
{Lovelace} R.~V.~E.,  {Li} H.,  {Colgate} S.~A.,   {Nelson} A.~F.,  1999,
  \mn@doi [\apj] {10.1086/306900}, \href
  {https://ui.adsabs.harvard.edu/abs/1999ApJ...513..805L} {513, 805}

\bibitem[\protect\citeauthoryear{{Miles}}{{Miles}}{1958}]{Miles1958}
{Miles} J.~W.,  1958, \mn@doi [Journal of Fluid Mechanics]
  {10.1017/S0022112058000653}, \href
  {https://ui.adsabs.harvard.edu/abs/1958JFM.....4..538M} {4, 538}

\bibitem[\protect\citeauthoryear{{Miranda} \& {Rafikov}}{{Miranda} \&
  {Rafikov}}{2019}]{Miranda2019}
{Miranda} R.,  {Rafikov} R.~R.,  2019, \mn@doi [\apjl]
  {10.3847/2041-8213/ab22a7}, \href
  {https://ui.adsabs.harvard.edu/abs/2019ApJ...878L...9M} {878, L9}

\bibitem[\protect\citeauthoryear{{Miranda} \& {Rafikov}}{{Miranda} \&
  {Rafikov}}{2020a}]{Miranda2020a}
{Miranda} R.,  {Rafikov} R.~R.,  2020a, \mn@doi [\apj]
  {10.3847/1538-4357/ab791a}, \href
  {https://ui.adsabs.harvard.edu/abs/2020ApJ...892...65M} {892, 65}

\bibitem[\protect\citeauthoryear{{Miranda} \& {Rafikov}}{{Miranda} \&
  {Rafikov}}{2020b}]{Miranda2020b}
{Miranda} R.,  {Rafikov} R.~R.,  2020b, \mn@doi [\apj]
  {10.3847/1538-4357/abbee7}, \href
  {https://ui.adsabs.harvard.edu/abs/2020ApJ...904..121M} {904, 121}

\bibitem[\protect\citeauthoryear{{Narayan} \& {Popham}}{{Narayan} \&
  {Popham}}{1993}]{Narayan&Popham1993}
{Narayan} R.,  {Popham} R.,  1993, \mn@doi [\nat] {10.1038/362820a0}, \href
  {https://ui.adsabs.harvard.edu/abs/1993Natur.362..820N} {362, 820}

\bibitem[\protect\citeauthoryear{{Narayan}, {Goldreich}  \&
  {Goodman}}{{Narayan} et~al.}{1987}]{Narayan+1987}
{Narayan} R.,  {Goldreich} P.,   {Goodman} J.,  1987, \mn@doi [\mnras]
  {10.1093/mnras/228.1.1}, \href
  {https://ui.adsabs.harvard.edu/abs/1987MNRAS.228....1N} {228, 1}

\bibitem[\protect\citeauthoryear{{Ono}, {Muto}, {Takeuchi}  \& {Nomura}}{{Ono}
  et~al.}{2016}]{Ono+2016}
{Ono} T.,  {Muto} T.,  {Takeuchi} T.,   {Nomura} H.,  2016, \mn@doi [\apj]
  {10.3847/0004-637X/823/2/84}, \href
  {https://ui.adsabs.harvard.edu/abs/2016ApJ...823...84O} {823, 84}

\bibitem[\protect\citeauthoryear{{Ono}, {Muto}, {Tomida}  \& {Zhu}}{{Ono}
  et~al.}{2018}]{Ono+2018}
{Ono} T.,  {Muto} T.,  {Tomida} K.,   {Zhu} Z.,  2018, \mn@doi [\apj]
  {10.3847/1538-4357/aad54d}, \href
  {https://ui.adsabs.harvard.edu/abs/2018ApJ...864...70O} {864, 70}

\bibitem[\protect\citeauthoryear{{Owen} \& {Menou}}{{Owen} \&
  {Menou}}{2016}]{Owen2016}
{Owen} J.~E.,  {Menou} K.,  2016, \mn@doi [\apjl]
  {10.3847/2041-8205/819/1/L14}, \href
  {https://ui.adsabs.harvard.edu/abs/2016ApJ...819L..14O} {819, L14}

\bibitem[\protect\citeauthoryear{{Papaloizou} \& {Pringle}}{{Papaloizou} \&
  {Pringle}}{1984}]{Papaloizou&Pringle1984}
{Papaloizou} J.~C.~B.,  {Pringle} J.~E.,  1984, \mn@doi [\mnras]
  {10.1093/mnras/208.4.721}, \href
  {https://ui.adsabs.harvard.edu/abs/1984MNRAS.208..721P} {208, 721}

\bibitem[\protect\citeauthoryear{{Pessah} \& {Chan}}{{Pessah} \&
  {Chan}}{2012}]{Pessah&Chan2012}
{Pessah} M.~E.,  {Chan} C.-k.,  2012, \mn@doi [\apj]
  {10.1088/0004-637X/751/1/48}, \href
  {https://ui.adsabs.harvard.edu/abs/2012ApJ...751...48P} {751, 48}

\bibitem[\protect\citeauthoryear{{Philippov}, {Rafikov}  \&
  {Stone}}{{Philippov} et~al.}{2016}]{Philippov+2016}
{Philippov} A.~A.,  {Rafikov} R.~R.,   {Stone} J.~M.,  2016, \mn@doi [\apj]
  {10.3847/0004-637X/817/1/62}, \href
  {https://ui.adsabs.harvard.edu/abs/2016ApJ...817...62P} {817, 62}

\bibitem[\protect\citeauthoryear{{Piro} \& {Bildsten}}{{Piro} \&
  {Bildsten}}{2004}]{Piro&Bildsten2004}
{Piro} A.~L.,  {Bildsten} L.,  2004, \mn@doi [\apj] {10.1086/421763}, \href
  {https://ui.adsabs.harvard.edu/abs/2004ApJ...610..977P} {610, 977}

\bibitem[\protect\citeauthoryear{{Popham} \& {Narayan}}{{Popham} \&
  {Narayan}}{1995}]{Popham&Narayan1995}
{Popham} R.,  {Narayan} R.,  1995, \mn@doi [\apj] {10.1086/175444}, \href
  {https://ui.adsabs.harvard.edu/abs/1995ApJ...442..337P} {442, 337}

\bibitem[\protect\citeauthoryear{{Popham}, {Narayan}, {Hartmann}  \&
  {Kenyon}}{{Popham} et~al.}{1993}]{Popham+1993}
{Popham} R.,  {Narayan} R.,  {Hartmann} L.,   {Kenyon} S.,  1993, \mn@doi
  [\apjl] {10.1086/187049}, \href
  {https://ui.adsabs.harvard.edu/abs/1993ApJ...415L.127P} {415, L127}

\bibitem[\protect\citeauthoryear{{Pringle}}{{Pringle}}{1977}]{Pringle1977}
{Pringle} J.~E.,  1977, \mn@doi [\mnras] {10.1093/mnras/178.2.195}, \href
  {https://ui.adsabs.harvard.edu/abs/1977MNRAS.178..195P} {178, 195}

\bibitem[\protect\citeauthoryear{{Revnivtsev} \& {Gilfanov}}{{Revnivtsev} \&
  {Gilfanov}}{2006}]{Revnivtsev&Gilfanov2006}
{Revnivtsev} M.~G.,  {Gilfanov} M.~R.,  2006, \mn@doi [\aap]
  {10.1051/0004-6361:20053964}, \href
  {https://ui.adsabs.harvard.edu/abs/2006A&A...453..253R} {453, 253}

\bibitem[\protect\citeauthoryear{{Shakura} \& {Sunyaev}}{{Shakura} \&
  {Sunyaev}}{1973}]{Shakura&Sunyaev1973}
{Shakura} N.~I.,  {Sunyaev} R.~A.,  1973, \aap, \href
  {https://ui.adsabs.harvard.edu/abs/1973A&A....24..337S} {24, 337}

\bibitem[\protect\citeauthoryear{{Stone}, {Tomida}, {White}  \&
  {Felker}}{{Stone} et~al.}{2020}]{Stone+2020}
{Stone} J.~M.,  {Tomida} K.,  {White} C.~J.,   {Felker} K.~G.,  2020, \mn@doi
  [\apjs] {10.3847/1538-4365/ab929b}, \href
  {https://ui.adsabs.harvard.edu/abs/2020ApJS..249....4S} {249, 4}

\bibitem[\protect\citeauthoryear{{Takasao}, {Hosokawa}, {Tomida}  \&
  {Iwasaki}}{{Takasao} et~al.}{2025}]{Takasao2025}
{Takasao} S.,  {Hosokawa} T.,  {Tomida} K.,   {Iwasaki} K.,  2025, \mn@doi
  [\apj] {10.3847/1538-4357/adc37b}, \href
  {https://ui.adsabs.harvard.edu/abs/2025ApJ...985...16T} {985, 16}

\bibitem[\protect\citeauthoryear{Virtanen et~al.,}{Virtanen
  et~al.}{2020}]{virtanen_scipy_2020}
Virtanen P.,  et~al., 2020, \mn@doi [Nature Methods]
  {10.1038/s41592-019-0686-2}, 17, 261

\bibitem[\protect\citeauthoryear{{Zhang} \& {Zhu}}{{Zhang} \&
  {Zhu}}{2020}]{Zhang2020}
{Zhang} S.,  {Zhu} Z.,  2020, \mn@doi [\mnras] {10.1093/mnras/staa404}, \href
  {https://ui.adsabs.harvard.edu/abs/2020MNRAS.493.2287Z} {493, 2287}

\bibitem[\protect\citeauthoryear{{Ziampras}, {Nelson}  \& {Rafikov}}{{Ziampras}
  et~al.}{2023}]{Ziampras2023}
{Ziampras} A.,  {Nelson} R.~P.,   {Rafikov} R.~R.,  2023, \mn@doi [\mnras]
  {10.1093/mnras/stad1973}, \href
  {https://ui.adsabs.harvard.edu/abs/2023MNRAS.524.3930Z} {524, 3930}

\makeatother
\end{thebibliography}

%\begin{thebibliography}{99}
%\end{thebibliography}

%%%%%%%%%%%%%%%%%%%%%%%%%%%%%%%%%%%%%%%%%%%%%%%%%%

%%%%%%%%%%%%%%%%% APPENDICES %%%%%%%%%%%%%%%%%%%%%

\appendix

%%%%%%%%%%%%%%%%%%%%%%%%%%%%%%%%%%%%
%%%%%%%%%%%%%%%%%%%%%%%%%%%%%%%%%%%%

\section{Smoothed Temperature Profile}
\label{app:k}

%%%%%%%%%%%%%%%%%%%%%%%%%%%%%%%%%%%%

In this appendix we detail the reasons behind the value of $k$ in  eq. \eqref{eq:T}. We start by noting that the value of $n$ has a large effect on the shape of $T_s$, with large values of $n$ producing flatter $T_s$ profiles. As a result, choosing one value of $k$, while straightforward, will lead to very different temperature profiles around the BL for different polytropic indices.

We therefore choose the value of $k$ that depends on $n$ to ensure a transition in the temperature profile which is similar across all values of $n$. To do this, first note that $T(1)=2^{1/k}T_0$. Given this, we then specify the temperature at some small distance $\delta r$ outside $1$ to be given by
\begin{equation}
    \label{eq:App_T_dr}
    T(1 + \delta r) = T_0 + f\times(T(1) - T_0)  ,
\end{equation}
where $f$ is a small numerical factor. The effect of this is to specify that the temperature a small distance $\delta R$ outside $1$, is a small factor $f$ of the temperature at $1$, where both temperatures are measured relative to $T_0$.

Substituting eqs. \eqref{eq:T_d}, \eqref{eq:T_s} and \eqref{eq:T} into \eqref{eq:App_T_dr}, we find
\begin{equation}
    \left(1 + \left[1-\frac{\gamma\mathcal{M}^2}{1+n}\left(1-\frac{1}{1+\delta r}\right)\right]^k\right)^{1/k} = 1 + f(2^{1/k} - 1)  .
\end{equation}
Assuming that $\delta r\ll 1$ (in our simulations, we use $\delta r = 0.01$) and rearranging gives
\begin{equation}
    1 + \left[1-\frac{\gamma\mathcal{M}^2}{1+n}\delta r\right]^k
    = \exp \left\{k\ln\left[1 + f(2^{1/k} - 1)\right] \right\}  .
\end{equation}
Further assuming that $f\ll1$,
\begin{equation}
    1 + \left[1-\frac{\gamma\mathcal{M}^2}{1+n}\delta r\right]^k
    = \exp \left\{fk(2^{1/k} - 1) \right\}  ,
\end{equation}
and finally that $k\gg1$ we arrive at
\begin{equation}
    1 + \left[1-\frac{\gamma\mathcal{M}^2}{1+n}\delta r\right]^k
    = e^{f\ln2}  ,
\end{equation}
which can be rearranged to give the desired expression for $k$:
\begin{equation}
    k = \frac{\ln\left(e^{f\ln2} - 1\right)}
    {\ln\left(1-\frac{\gamma\mathcal{M}^2}{1+n}{\delta r}\right)}  .
        \label{eq:k_value}
\end{equation}
All the simulations in this work use $\gamma=1.00001$, ${f=0.1}$ and ${\delta r=0.01}$. 

%%%%%%%%%%%%%%%%%%%%%%%%%%%%%%%%%%%%
%%%%%%%%%%%%%%%%%%%%%%%%%%%%%%%%%%%%

\section{Simulation Outputs} 
\label{sec:outputs}

%%%%%%%%%%%%%%%%%%%%%%%%%%%%%%%%%%%%

The entire state of the simulation is output on a cadence of once an orbit, giving a total of 1000 outputs for each simulation. In addition to this, several quantities are output at a much higher cadence to allow for more detailed analysis. These higher cadence files are output 2x10 times per orbit, one set spaced by 0.1 orbits and the other set spaced by the same amount but offset by 0.01 orbits (i.e. outputs at 0.00, 0.01, 0.10, 0.11 etc. of an orbit). The reason for doing this cadence is to allow us to capture temporal behaviour on a timescale of ${t/2\pi=0.01}$ without requiring 100 outputs per orbit.

One set of the high cadence files consists of various azimuthally averaged quantities. This gives us the radial profiles of quantities such as ${\langle\Sigma\rangle_\phi(r)}$, ${\langle v_\phi\rangle_\phi(r)}$, ${\langle\Sigma v_\phi\rangle_\phi(r)}$ etc. Additionally, we have azimuthal profiles from two different radii, one at $r=1.1$ and one halfway between the inner edge of the domain (see Table \ref{tab:all_sims}) and $r=1$. These files are used to identify the different modes present in the star and disc in each simulation.

%%%%%%%%%%%%%%%%%%%%%%%%%%%%%%%%%%%%
%%%%%%%%%%%%%%%%%%%%%%%%%%%%%%%%%%%%

\section{Mode detection procedure} 
\label{sec:mode-detect}

%%%%%%%%%%%%%%%%%%%%%%%%%%%%%%%%%%%%

Modes are identified based on the stability of their pattern speed $\OmegaP$, independently of the mode strength (although weaker modes will be affected my noise more greatly, and so implicitly we expect to be biased towards stronger modes). To do this, ollowing the mode detection procedure used in \citet{Coleman+2022a,Coleman+2022b}, the simulation is divided into ten orbits chunks. Each chunk contains 100 pairs of files, with an intra-pair spacing of ${t/2\pi=0.01}$ and an inter-pair spacing of ${t/2\pi=0.1}$. For each pair of files, the pattern speeds ${\OmegaP(m)}$ is calculated, by considering the phase shift in the mode over ${t/2\pi=0.01}$. This gives 100 sets of ${\OmegaP(m)}$ for each ten orbit chunk.

Identifying which modes are present is a two-step process. First, for each value of $m$ up to 40, a linear fit is performed to ${\OmegaP(m,t)}$ against time. We use three different levels of threshold for this initial identification. The most stringent threshold requires that the evolution of $\OmegaP$ in each orbit is limited to $|2\pi{\mathrm{d}\OmegaP}/{\mathrm{d}t}|<(2\times10^{-4}, 4\times10^{-4})$ where the different threshold apply to the star and disc respectively. Additionally, the residual standard deviation from the fit must be less than $(1\times10^{-4}, 2\times10^{-4})$. The other two detection thresholds are similar, but threshold is a factor of three or ten larger. A mode at the highest threshold is given a `quality' of three, with the lower threshold being of quality two and one. 

Once this has been performed, we identify each individual mode, defined as a period of one or more consecutive chunks (each chunk being ten orbits long) where the mode meets at least the lowest detection threshold. For each period, we sum the modes qualities from each chunk. Again, this summation has three different detection thresholds of three, six and nine to count as a detected mode. For example, to reach the lowest of these threshold, a mode could have one chunk at the highest threshold, or three consecutive chunks at the lowest threshold. While this process seems unnecessarily complicated, it was designed to ensure that both high quality, short-lived modes and lower quality but longer-lived modes are detected at the highest threshold. Indeed, there are examples of modes meeting the highest threshold by virtue of being identified at the lowest quality in nine or more consecutive chunks.

\section{Analytic dispersion relations}
\label{app:mode_eqs}

The analytic dispersion relation for lower modes is given by equation (39) of \citet{Belyaev+2013a} which is reproduced here:
\begin{equation}
    \label{eq:lower_analytic}
    \OmegaP = \frac{1}{r_0}\sqrt{\mathcal{M}^{-2} + \left(\frac{\mathcal{M}}{2mr_0}\right)^2} .
\end{equation}
The parameter $r_0$ is a free parameter which was chosen to provide a best-fit by eye to the modes present in Figure \ref{fig:all_modes}.

Similarly, the relation for upper modes is given by equation (A2) of \citet{Coleman+2022a},
\begin{equation}
    \label{eq:upper_analytic}
    m^2 = \frac{\kappa^2(\rf)\rf^2\left[\Omega(\rf)-\OmegaP\right] - 2c_s^2\Omega(\rf)}
    {\rf^2\left[\Omega(\rf)-\OmegaP\right]^3 - c_s^2\left[\Omega(\rf)-\OmegaP\right]} ,
\end{equation}
where $\kappa$ is the radial epicyclic frequency, and $\rf$ is a fiducial radius, given by the location of the maximum of the angular velocity profile. The $\Omega$ and $\kappa$ values required were calculated as averages across the full time domain (excluding the first 20 orbits) of each simulation.

%%%%%%%%%%%%%%%%%%%%%%%%%%%%%%%%%%%%%%%%%%%%%%%%%%

% Don't change these lines
\bsp	% typesetting comment
\label{lastpage}
\end{document}